\documentclass[prb,aps,superscriptaddress,twocolumn,dvips]{revtex4}
\usepackage{feynmp}
\usepackage{amsmath}
\usepackage{amssymb}
\usepackage{epsfig}
\usepackage{graphicx}
\usepackage{subfigure}
\usepackage{hyperref}
\usepackage{bbm}
\usepackage{color}
\usepackage{longtable}
\usepackage{booktabs}
\usepackage{multirow}
\newcommand{\tabincell}[2]{\begin{tabular}{@{}#1@{}}#2\end{tabular}}

\newcommand{\Rmnum}[1]{\expandafter\@slowromancap\romannumeral #1@}
\allowdisplaybreaks[3]
\begin{document}

\title{Quantum phase transition and unusual critical behavior in multi-Weyl semimetals}

\author{Jing-Rong Wang}
\affiliation{Anhui Province Key Laboratory of Condensed Matter
Physics at Extreme Conditions, High Magnetic Field Laboratory of the
Chinese Academy of Science,Hefei 230031, Anhui, China}
\author{Guo-Zhu Liu}
\altaffiliation{Corresponding author: gzliu@ustc.edu.cn}
\affiliation{Department of Modern Physics, University of Science and
Technology of China, Hefei, Anhui 230026, P. R. China}
\author{Chang-Jin Zhang}
\altaffiliation{Corresponding author: zhangcj@hmfl.ac.cn}
\affiliation{Anhui Province Key Laboratory of Condensed Matter
Physics at Extreme Conditions, High Magnetic Field Laboratory of the
Chinese Academy of Science,Hefei 230031, Anhui,
China}\affiliation{Collaborative Innovation Center of Advanced
Microstructures, Nanjing University, Nanjing 210093, P. R. China}

\begin{abstract}
The low-energy behaviors of gapless double- and triple-Weyl fermions
caused by the interplay of long-range Coulomb interaction and
quenched disorder are studied by performing a renormalization group
analysis. It is found that an arbitrarily weak disorder drives the
double-Weyl semimetal to undergo a quantum phase transition into a
compressible diffusive metal, independent of the disorder type and
the Coulomb interaction strength. In contrast, the nature of the
ground state of triple-Weyl fermion system relies sensitively on the
specific disorder type in the noninteracting limit: The system is
turned into a compressible diffusive metal state by an arbitrarily
weak random scalar potential or $z$ component of random vector
potential but exhibits stable critical behavior when there is only
$x$ or $y$ component of random vector potential. In case the
triple-Weyl fermions couple to random scalar potential, the system
becomes a diffusive metal in the weak interaction regime but
remains a semimetal if Coulomb interaction is sufficiently strong.
Interplay of Coulomb interaction and $x$, or $y$, component of
random vector potential leads to a stable infrared fixed point that
is likely to be characterized by critical behavior. When Coulomb
interaction coexists with the $z$ component of random vector potential,
the system flows to the interaction-dominated strong coupling
regime, which might drive a Mott insulating transition. It is thus
clear that double- and triple-Weyl fermions exhibit distinct
low-energy behavior in response to interaction and disorder. The
physical explanation of such distinction is discussed in detail. The
role played by long-range Coulomb impurity in triple-Weyl semimetal
is also considered. The main conclusion is that, Coulomb impurity
always drives the system to become a compressible diffusive metal,
whereas Coulomb interaction tends to suppress the Coulomb impurity,
rendering the robustness of the semimetal phase.
\end{abstract}

\maketitle


\section{Introduction}

Topological semimetals (SMs), including Dirac semimetal (DSM)
\cite{Vafek14, Wehling14, Armitage17}, Weyl semimetal (WSM)
\cite{Armitage17, Wan11, Burkov16, Yan17, Hasan17}, and nodal line
semimetal (NLSM) \cite{Armitage17, Weng16, FangChen16}, have
attracted broad research interest among the condensed-matter
community. These materials exhibit many intriguing properties,
provide ideal platforms for exploring some important physical
concepts, and also have promising industrial applications.
Specifically, WSM has been theoretically predicted and
experimentally observed in the TaAs family \cite{Huang15, WengHM15,
Xu15A, Lv15A}. In usual WSMs, the fermion excitations emerge at low
energies from pairs of Weyl nodes with opposite monopole charges
$\pm 1$, and have a linear-in-momentum dispersion \cite{Wan11,
Burkov16, Yan17}. They are known to host symmetry-protected Fermi
arc surface states \cite{Wan11, Burkov16, Yan17} and display chiral
anomaly that is related to the presence of negative
magnetoresistance \cite{HuangChenGFGroup15, ZhangChengLong16}. Apart
from usual WSMs, there are also multi-WSMs in which the monopole
charges of Weyl nodes can be larger than unity \cite{XuFang11,
Fang12, YangNogasa14, Bradlyn16, Gao16, Huang16, Guan15,
LiuZunger17, ChenChan16, ChangChan17, Lai15, Jian15, Zhang16,
WangLiuZhang16, Jian16, Goswami15, Roy15, BeraRoy16, LiRoy16, Roy17,
Ahn16, Ahn17, Park17, Huang17, Hayata17, Sun17, Shapourian16, Sbierski17}.
Two known examples are double- and triple-WSMs, where the fermion
dispersion is linear in one of the momentum components, but
quadratic and cubic in the other two components. Accordingly, the
monopole charges of double- and triple-Weyl fermions are $\pm 2$ and
$\pm 3$, respectively.

Differen from ordinary metals that possess a finite Fermi surface,
multi-WSMs have only zero-dimensional Fermi points. Accordingly, the
density of states (DOS) vanishes at zero energy, i.e. $\rho(0)=0$,
which renders the robustness of the long-range character of Coulomb
interaction. The dynamics of multi-Weyl fermions are distinct from
ordinary electrons when they are surrounded by a randomly
distributed potential. The impact of Coulomb interaction
\cite{Shankar94, ColemanBook, Varma02, Kotov12} and disorder
\cite{Lee85, Evers08, DasSarma11, Syzranov16Review} on the
low-energy behavior of multi-WSMs might be drastically different
from normal metals. The purpose of the present paper is to give a
comprehensive theoretic analysis of the low-energy behavior of
double- and triple-WSMs in the presence of both Coulomb interaction
and disorder.

\begin{table*}[htbp]
\caption{Summary of the possible ground states of double- and triple-WSMs
due to the interplay between Coulomb interaction and disorder.
Here, $\alpha_0$ is the bare strength parameter for Coulomb
interaction. We use $\Delta_{00}$, $\Delta_{10}$, $\Delta_{20}$, and
$\Delta_{30}$ to denote the bare disorder parameter for RSP,
$x$-RVP, $y$-RVP, and $z$-RVP, respectively. Morever, CDM stands for
compressive diffusive metal, SM for semimetal, and SCS for stable
critical state. \label{Table:MainResults}}
\begin{center}
\begin{tabular}{|c|c|c|c|c|c|}
\hline\hline  \multirow{2}{*}{} & \multicolumn{2}{|c|}{Double-Wely
Semimetal} & \multicolumn{3}{|c|}{Triple-Weyl Semimetal}
\\ \cline {2-6} & $\alpha_{0}=0$ & $\alpha_{0}>0$
& $\alpha_{0}=0$ & \multicolumn{2}{|c|}{$\alpha_{0}>0 $}
\\
\hline \multirow{2}{*}{$\Delta_{00}>0$} &
\multirow{5}{*}{\tabincell{c}{$\Delta_{0}\rightarrow\infty$\\CDM}} &
\multirow{5}{*}{
\tabincell{c}{$\Delta_{0}\rightarrow\infty,\alpha\rightarrow\infty,
\alpha/\Delta_{0}\rightarrow0$\\CDM}} &
\multirow{2}{*}{\tabincell{c}{$\Delta_{0}\rightarrow\infty$ \\CDM}}
& $\alpha_{0}<\alpha_{0c}(\Delta_{00})$ &
$\alpha_{0}>\alpha_{0c}(\Delta_{00})$ \\ \cline{5-6} & & & &
\tabincell{c}{$\alpha\rightarrow\infty$,
$\Delta_{0}\rightarrow\infty$, $\alpha/\Delta_{0}\rightarrow0$\\CDM
}& \tabincell{c}{$\alpha\rightarrow0$, $\Delta_{0}\rightarrow0$,
$\Delta_{0}/\alpha\rightarrow0$\\SM }  \\ \cline{1-1} \cline{4-6}
$\Delta_{10}>0$ & &  &
\tabincell{c}{$\Delta_{1}\rightarrow\Delta_{1}^{*}$\\SCS} &
\multicolumn{2}{|c|}{\tabincell{c}{$\Delta_{1}\rightarrow\Delta_{1n}^{*},
\alpha\rightarrow\alpha_{n}^{*}$\\ Possible SCS}}
\\ \cline{1-1} \cline{4-6}  $\Delta_{20}>0$ & & &
\tabincell{c}{$\Delta_{2}\rightarrow\Delta_{2}^{*}$\\SCS} &
\multicolumn{2}{|c|}{\tabincell{c}{$\Delta_{2}\rightarrow\Delta_{2n}^{*},
\alpha\rightarrow\alpha_{n}^{*}$\\Possible SCS}}
\\ \cline{1-1} \cline{4-6} $\Delta_{30}>0$ & & &
\tabincell{c}{$\Delta_{3}\rightarrow\infty$ \\ CDM} &
\multicolumn{2}{|c|}{\tabincell{c}{ $\Delta_{3}\rightarrow\infty,
\alpha\rightarrow\infty, \alpha/\Delta_{3}>1$\\ Possible Mott
Insulator}}
\\
\hline\hline
\end{tabular}
\end{center}
\end{table*}

Previous renormalization group (RG) studies \cite{Lai15, Jian15}
showed that the Coulomb interaction in clean double-WSMs is
marginally irrelevant, which results in logarithmic-like corrections
for some observable quantities. The RG analysis of Zhang \emph{et
al.} \cite{Zhang16} found that the Coulomb interaction is also
marginally irrelevant in clean triple-WSMs. In a recent work
\cite{WangLiuZhang16}, we carried out a RG study of the
quasiparticle residue $Z_f$ and the Landau damping rate after
including the energy dependence of dressed Coulomb interaction, and
illustrated that the conventional Fermi liquid (FL) description is
spoiled in both double- and triple-WSMs in an anomalous manner. In
particular, though the Coulomb interaction only leads to a
relatively weak Landau damping effect, the residue $Z_f$ vanishes in
the lowest energy limit, resulting in a new type of non-Fermi liquid
(NFL) state. In such a NFL state, the FL theory is violated more
weakly than a marginal Fermi liquid (MFL), which has long been
regarded as the weakest breakdown of FL theory \cite{Varma02}.

The disorder effects on non-interacting double-Weyl fermions have
been studied by means of several methods, including self-consistent
Born approximation (SCBA) \cite{Goswami15}, RG approach
\cite{BeraRoy16}, and large-scale quantum simulations
\cite{Sbierski17}. It was showed \cite{BeraRoy16, Syzranov16Review, Fradkin86,
Goswami11, Kobayashi14, Sbierski14, Roy14, Syzranov15A, Syzranov15B,
Pixley15, Sbierski15, Roy16Erratum, Liu16, Roy16A,
Syzranov16A, Louvet16, Pixley16A, Pixley16B, Pixley16C, Roy16B,
Roy16C, Sbierski16, Fu17, Pixley17} that even an arbitrarily weak
disorder is able to drive the double-WSM to enter into a
compressible diffusive metal (CDM) phase (see, however,
Ref.~\cite{Shapourian16}), which is characterized by the generation
of finite zero-energy disorder scatting rate $\gamma_{0}$ and finite
zero-energy DOS $\rho(0)$. One might regard $\rho(0)$ as an
effective order parameter for the CDM phase. Moreover, the electric
conductivity is finite at zero temperature in this phase. A naive
power counting suggests that short-range disorder is a relevant
perturbation in a triple-WSM \cite{BeraRoy16}. However, a detailed
RG analysis of disorder effects on triple-Weyl fermions is still
lacking.

When both Coulomb interaction and disorder are considered, their
interplay might give rise to a variety of intriguing properties
\cite{Finkelstein84, Castellani84, PunnooseFinkelstein05,
Abrahams01, Kravchenko04, Spivak10, Ye98, Ye99, Stauber05, Herbut08,
Vafek08, Foster08, WangLiu14, Zhao16, Goswami11, Roy16A, Gonzalez17,
Moon14, Nandkishore17, YuXuanWang17}. This problem has been
extensively studied for over three decades, and plays a vital role
in the studies of two-dimensional (2D) metallic systems
\cite{Finkelstein84, Castellani84, PunnooseFinkelstein05,
Abrahams01, Kravchenko04, Spivak10}. It is believed by many
researchers that the observed metal-insulator transition in some 2D
systems is driven by an intricate interplay of strong Coulomb
interaction and disorder scattering \cite{Finkelstein84,
Castellani84, PunnooseFinkelstein05, Abrahams01}. Unfortunately, the
progress of this subject is extremely slow, and there is still not
an unified framework that treats the interplay of interaction and
disorder in a satisfactory way.

Because of the peculiar geometry of Fermi surface of SMs, the
interplay of Coulomb interaction and disorder may lead to distinct
low-energy behavior comparing to traditional metals \cite{Ye98,
Ye99, Stauber05, Herbut08, Vafek08, Foster08, Goswami11, WangLiu14,
Zhao16, Roy16A, Gonzalez17, Moon14, Nandkishore17, YuXuanWang17}. In
this paper, we present a systematic RG study of the interplay of
Coulomb interaction and quenched disorder, including random scalar
potential (RSP) and random vector potential (RVP), in both double-
and triple-WSMs. Normally, the Coulomb interaction tends to suppress
the low-energy DOS of Dirac/Weyl fermions \cite{Vafek14, Wehling14,
Kotov12}. In contrast, disorder can usually enhance the fermion DOS
\cite{Kotov12, Evers08, DasSarma11, Syzranov16Review}. The true
low-energy dynamics of the fermions should be determined by an
appropriate treatment of the mutual influence between interaction
and disorder \cite{Ye98, Ye99, Stauber05, Herbut08, Vafek08,
Foster08, Goswami11, WangLiu14, Moon14, Zhao16, Roy16A, Gonzalez17,
Nandkishore17, YuXuanWang17}. Our RG analysis reveals a remarkable
difference between double- and triple-WSMs. In particular, we find
that, an arbitrarily weak disorder is able to turn double-WSM into a
CDM. It is important to notice that such a transition happens for
any type of disordered potential and is independent of the Coulomb
interaction strength. In sharp contrast to the double-WSM, the
subtle interplay between Coulomb interaction and disorder can give
rise to a variety of different ground states and various quantum
phase transitions (QPT) in a triple-WSM, which is summarized in
Table~\ref{Table:MainResults}. In the non-interacting limit, the
system is turned into a CDM phase by an arbitrarily weak RSP or
$z$-component of RVP, dubbed $z$-RVP hereafter, but exhibits stable
critical state when there is only $x$-RVP or $y$-RVP. In case the
triple-Weyl fermions couple to RSP, the system becomes a CDM in the
weak interaction regime, but is turned back to a SM if the Coulomb
interaction becomes sufficiently strong. The interplay of Coulomb
interaction and $x$-RVP, or $y$-RVP, produces a stable fixed point
that is likely to be characterized by the emergence of unusual
critical behavior. When Coulomb interaction coexists with $z$-RVP,
the system flows to the interaction-dominated strong coupling
regime, which might drive a Mott insulating transition. If two or
more types of disorder coexist, triple-WSM is always driven to
undergo a CDM transition, no matter the Coulomb interaction is
included or not, which is similar to double-WSM.

All the above considerations are restricted to short-range disorder.
In various SMs, there might be disorder with long-range correlation,
such as Coulomb impurity. The long-range disorder may be more
important than short-range disorder in SM systems. We will also
apply the RG approach to study the interplay of long-range Coulomb
interaction, long-range Coulomb impurity, and short-range RSP in
triple-WSM. Our main conclusion is that, the Coulomb impurity can
always drive a SM-to-CDM phase transition, whereas the Coulomb
interaction tends to suppress the Coulomb impurity, rendering the
robustness of the SM phase.

The rest of the paper is structured as follows. We present the
effective action in Sec.~\ref{Sec:Model}, and discuss the RG results
in Sec.~\ref{Sec:Results}. We address several related issues in
Sec~\ref{Sec:Discussions}. The results are briefly summarized in
Sec.~\ref{Sec:Summary}. The detailed derivations of the RG equations
are presented in the Appendices.

\section{Effective action\label{Sec:Model}}

The free Hamiltonian for double-Weyl fermions is \cite{Lai15,
Jian15, Jian16, WangLiuZhang16, Zhang16}
\begin{eqnarray}
H_{d} = \int d^3\mathbf{x}
\psi_{d}^{\dag}(\mathbf{x})\left[Ad_{i}(\mathbf{x})\sigma_{i} -
iv\partial_{z}\sigma_{3}\right]\psi_{d}(\mathbf{x}),
\end{eqnarray}
where $d_{1}(\mathbf{x})=-\left(\partial_{x}^{2} -
\partial_{y}^{2}\right)$ and $d_{2}(\mathbf{x}) =
- 2\partial_{x}\partial_{y}$. The free Hamiltonian for triple-Weyl
fermions is \cite{Zhang16, WangLiuZhang16}
\begin{eqnarray}
H_{t} = \int d^3\mathbf{x} \psi_{t}^{\dag}(\mathbf{x})
\left[Bg_{i}(\mathbf{x})\sigma_{i} -
iv\partial_{z}\sigma_{3}\right]\psi_{t}(\mathbf{x}),
\end{eqnarray}
where $g_{1}(\mathbf{x}) = i\left(\partial_{x}^{3} -
\partial_{x}\partial_{y}^{2}\right)$ and
$g_{2}(\mathbf{x}) = i\left(\partial_{y}^{3} -
\partial_{y}\partial_{x}^{2}\right)$. Here, we use $\psi_{d}$ and
$\psi_{t}$ to represent two-component spinor for double- and
triple-Weyl fermions, respectively. Moreover, $\sigma_{1, 2, 3}$ are
the Pauli matrices. The dispersions of double- and triple-Weyl
fermions are defined as
\begin{eqnarray}
E_{\pm}^{d}(k) &=& \pm\sqrt{A^{2}k_{\bot}^{4}+v^{2}k_{z}^{2}},\\
E_{\pm}^{t}(k) &=& \pm\sqrt{B^{2}k_{\bot}^{6} + v^2k_{z}^{2}},
\end{eqnarray}
where $A$, $B$, and $v$ are model parameters. The long-range Coulomb
interaction between fermions can be written as
\begin{eqnarray}
H_{\mathrm{C}} = \frac{1}{4\pi}\int d^3\mathbf{x} d^3
\mathbf{x}'\rho_{d,t}(\mathbf{x}) \frac{e^2}{\epsilon\left|\mathbf{x} -
\mathbf{x}'\right|}\rho_{d,t}^{\dag}(\mathbf{x}'),
\end{eqnarray}
where $\rho_{d,t}(\mathbf{x}) = \psi_{d,t}^{\dag}(\mathbf{x})
\psi_{d,t}(\mathbf{x})$ is the fermion density operator, $e$ is the
electric charge, and $\epsilon$ is the dielectric constant. The
effective strength of Coulomb interaction is defined as $\alpha =
e^{2}/v\epsilon$. The action for fermion-disorder coupling reads
\begin{eqnarray}
S_{\mathrm{dis}} = \sum_{j=0}^{3}\int d\tau d^3\mathbf{x}V_{j}
\psi_{d,t}^{\dag}\Gamma_{j}\psi_{d,t}.
\end{eqnarray}
The quenched random field $V_{j}$ is taken as a Gaussian white noise
distribution that satisfies $\langle V_{j}(\mathbf{x})\rangle = 0$
and $\langle V_{j}(\mathbf{x})V_{j}(\mathbf{x}')\rangle =
\Delta_{j}\delta^{3}(\mathbf{x}-\mathbf{x}')$. The Coulomb
interaction can be decoupled by introducing a bosonic field $\phi$
through Hubbard-Stratonovich transformation, whereas the disorder
can be treated by the replica method.

Now, the total effective action can be written as
\begin{eqnarray}
S&=&S_{f}+S_{b}+S_{fb},
\\
S_{f}&=&\int\frac{d\omega}{2\pi}\frac{d^3\mathbf{k}}{(2\pi)^{3}}
\psi_{a}^{\dag} \left[-i\omega+H_{f}(\mathbf{k})\right]\psi_{a},
\\
S_{b}&=&\int\frac{d\omega}{2\pi}\frac{d^3\mathbf{k}}{(2\pi)^{3}}\phi
\left(k_{x}^{2} + k_{y}^{2} + \eta k_{z}^{2}\right)\phi,
\\
S_{fb}&=&ig\int d\tau d^3\mathbf{x}\phi\psi_{a}^{\dag}\psi_{a},
\\
S_{\mathrm{dis}}&=&\sum_{j=0}^{3}\frac{\Delta_{j}}{2}\int d\tau
d\tau'd^3\mathbf{x} \left(\psi_{a}^{\dag} \Gamma_{j}
\psi_{a}\right)_{\tau}\left(\psi_{b}\Gamma_{j}\psi_{b}\right)_{\tau'},
\end{eqnarray}
where $g=\frac{\sqrt{4\pi}e}{\sqrt{\epsilon}}$ and $\eta$ is
introduced to parameterize the anisotropy of the Coulomb
interaction. The disordered potential is averaged by employing the
standard replica trick, with $a,b = 1,2,..,n$ being the replica
indices. At the end of the calculation, the limit $n\rightarrow 0$
will be taken. The disorder type is determined by the concrete
expression of the $\Gamma_{j}$ matrix. For $\Gamma_{0}=\mathbbm{1}$,
disorder plays the role of a RSP. The matrices
$\Gamma_{1,2,3}=\sigma_{1,2,3}$ correspond to the three components
of RVP. The effective strength of disorder is represented by
$\Delta_{j}$ with $j=0,1,2,3$. To analyze the impact of Coulomb
interaction and disorder, we need to derive the RG equations for all
the model parameters.

\section{Renormalization group results \label{Sec:Results}}

In this section, we make a systematic RG analysis. We have carried
out analytical calculations by making perturbative expansion in
powers of small coupling constants, with details presented in the
Appendices, and then derived the coupled RG equations for all the
model parameters in the cases of double- and triple-WSMs. For
simplicity, here we only write down the RG equations obtained in the
case of triple-WSM:
\begin{eqnarray}
\frac{dZ_{f}}{d\ell}&=&-\frac{5}{4}\sum_{j=0}^{3}\Delta_{j}Z_{f},
\\
\frac{dB}{d\ell}&=&\left(C_{2}^{t} -
\frac{5}{4}\sum_{j=0}^{3}\Delta_{j}\right)B,
\\
\frac{dv}{d\ell}&=&\left(C_{3}^{t} -
\frac{5}{4}\sum_{j=0}^{3}\Delta_{j}\right)v,
\\
\frac{d\alpha}{d\ell} &=&\left(-C_{\bot}^{t} - C_{3}^{t} +
\frac{5}{4}\sum_{j=0}^{3}\Delta_{j}\right)\alpha,
\\
\frac{d\beta^{t}}{d\ell} &=& \left(\frac{4}{3} -
\frac{2}{3}C_{2}^{t} +C_{3}^{t}- \beta^{t} - \frac{5}{12}
\sum_{0}^{3}\Delta_{j}\right)\beta^{t},
\\
\frac{d\eta}{d\ell}&=&\left(-\frac{4}{3} -
C_{\bot}^{t}+\beta^{t}\right)\eta,
\\
\frac{d\Delta_{0}}{d\ell} &=&\frac{1}{3}\Delta_{0} +
\left(\frac{25}{12}\Delta_{0}^{2} +
\frac{25}{12}\Delta_{0}\Delta_{1} +
\frac{25}{12}\Delta_{0}\Delta_{2}\right.\nonumber
\\
&&\left.+\frac{25}{12}\Delta_{0}\Delta_{3} +
3\Delta_{1}\Delta_{2}+\frac{1}{2}\Delta_{1}\Delta_{3}+
\frac{1}{2}\Delta_{2}\Delta_{3}\right)\nonumber
\\
&&-\Delta_{0}\left(\frac{2}{3}C_{2}^{t}+C_{3}^{t} +
2C_{\bot}^{t}+2\beta^{t}\right),
\\
\frac{d\Delta_{1}}{d\ell} &=&\frac{1}{3}\Delta_{1}+
\left(-\frac{23}{12}\Delta_{1}\Delta_{0} -
\frac{23}{12}\Delta_{1}^{2} +
\frac{13}{12}\Delta_{1}\Delta_{2}\right.\nonumber
\\
&&\left.+\frac{13}{12}\Delta_{1}\Delta_{3} + 3\Delta_{0}\Delta_{2} +
\frac{1}{2}\Delta_{0}\Delta_{3}\right)\nonumber
\\
&&+\Delta_{1}\left(2C_{4}^{t}-\frac{2}{3}C_{2}^{t}-C_{3}^{t}\right),
\\
\frac{d\Delta_{2}}{d\ell}
&=&\frac{1}{3}\Delta_{2}+\left(-\frac{23}{12}\Delta_{2}\Delta_{0} +
\frac{13}{12}\Delta_{2}\Delta_{1} -
\frac{23}{12}\Delta_{2}^{2}\right.\nonumber
\\
&&\left.+\frac{13}{12}\Delta_{2}\Delta_{3}+3\Delta_{0}\Delta_{1} +
\frac{1}{2}\Delta_{0}\Delta_{3}\right)\nonumber
\\
&&+\Delta_{2}\left(2C_{4}^{t}-\frac{2}{3}C_{2}^{t}-C_{3}^{t}\right),
\\
\frac{d\Delta_{3}}{d\ell} &=& \frac{1}{3}\Delta_{3} +
\left(\frac{1}{12}\Delta_{3}\Delta_{0} -
\frac{11}{12}\Delta_{3}\Delta_{1} -
\frac{11}{12}\Delta_{3}\Delta_{2}\right.\nonumber
\\
&&\left.+\frac{1}{12}\Delta_{3}^{2} + \Delta_{0}\Delta_{1} +
\Delta_{0}\Delta_{2}\right) - \Delta_{3}\left(\frac{2}{3}C_{2}^{t} -
C_{3}^{t}\right). \nonumber \\
\end{eqnarray}
In the derivation of these equations, we have made the re-definition
$\frac{c_{f}\Delta_{i}}{vB^{\frac{2}{3}}\Lambda^{\frac{1}{3}}}
\rightarrow \Delta_{i}$ with $c_{f} = \frac{\Gamma(\frac{1}{3})}{15
\pi^{\frac{3}{2}}\Gamma(\frac{5}{6})}$. We use $\ell$ to represent
the varying length scale and $Z_{f}$ to represent the quasiparticle
residue. The influence of the Coulomb interaction is encoded in the
functions $C_{i}^{t}\equiv C_{i}^{t}(\alpha,\zeta^{t})$,
$C_{\bot}^{t}$, and $\beta^{t}$, where $\zeta^{t}=\frac{\eta
B^{\frac{2}{3}}\Lambda^{\frac{4}{3}}}{v^2}$,
$C_{\bot}^{t}=\frac{\alpha}{\pi}$, and $\beta^{t}=\frac{\pi
c_{f}}{2}\frac{\alpha v^{2}}{B^{\frac{2}{3}}
\Lambda^{\frac{4}{3}}\eta}$. By taking $\alpha=0$, we can easily
obtain the RG equations for the case in which fermions couple solely
to disorder.

\subsection{Non-interacting limit}

As the first step, we consider the non-interacting limit by taking
$\alpha = 0$, and analyze the properties of double- and triple-WSMs
induced by disorder.

For double-WSM, there is an interesting correlation among RSP,
$x$-RVP, and $y$-RVP: if any of them exists solely with a finite
strength, the other two can be dynamically generated and all of the
three disorder parameters flow eventually to the strong coupling
regime. In the spirit of RG theory, the divergence of the strength
of some interaction in the lowest energy limit usually indicates
that this interaction is relevant and can lead to an instability of
the system \cite{Shankar94}. The divergence of disorder parameter is
often interpreted as the transition of the system into a CDM phase
that is characterized by the generation of a finite zero-energy DOS
\cite{ BeraRoy16, Syzranov16Review, Fradkin86, Goswami11, Kobayashi14,
Sbierski14, Roy14, Syzranov15A, Syzranov15B, Pixley15, Sbierski15,
Roy16Erratum, Liu16, Roy16A, Syzranov16A, Louvet16,
Pixley16A, Pixley16B, Pixley16C, Roy16B, Roy16C, Sbierski16, Fu17,
Pixley17}. Different from the other types of random potential,
$z$-RVP is always fixed at zero. However, in case the system
contains only $z$-RVP with a finite initial value, the other three
types of disorder would be dynamically generated, and all of the
four disorder parameters flow to the strong coupling regime. These
results indicate that $(\Delta_{0}^{*}, \Delta_{1}^{*}, \Delta_{2}^{*},
\Delta_{3}^{*}) = (0,0,0,0)$ is an unstable infrared fixed point, and
that an arbitrarily weak disorder is able to drive the system to
become a CDM, which are well consistent with Bera \emph{et al}
\cite{BeraRoy16}.

We now discuss the case of triple-WSM. If RSP exists by itself in
the triple-WSM, the RG equation for $\Delta_0$ is
\begin{eqnarray}
\frac{d\Delta_{0}}{d\ell} = \frac{1}{3}\Delta_{0} +
\frac{25}{12}\Delta_{0}^{2}.
\end{eqnarray}
Its solution is given by
\begin{eqnarray}
\Delta_{0} = \frac{\frac{4}{25}
e^{\frac{1}{3}\ell}\Delta_{00}}{\Delta_{00} + \frac{4}{25} -
e^{\frac{1}{3}\ell}\Delta_{00}}.
\end{eqnarray}
Obviously, there is only one infrared fixed point $\Delta_{0}^{*} =
0$, which is unstable, implying that any weak RSP induces a QPT from
triple-WSM to CDM.

If $x$-RVP exists alone, the RG equation satisfies
\begin{eqnarray}
\frac{d\Delta_{1}}{d\ell} = \frac{1}{3}\Delta_{1} -
\frac{23}{12}\Delta_{1}^{2},
\end{eqnarray}
which has the following solution
\begin{eqnarray}
\Delta_{1} = \frac{\frac{4}{23}e^{\frac{1}{3}\ell}
\Delta_{10}}{e^{\frac{1}{3}\ell}\Delta_{10} - \Delta_{10} +
\frac{4}{23}}.
\end{eqnarray}
We find that $\Delta_{1}^{*} = \frac{4}{23}$ is a stable infrared
fixed point. At low energies, the quasiparticle residue behaves as
\begin{eqnarray}
Z_{f}\sim e^{-\frac{5}{23}\ell},\label{Eq:ZfTWFP}
\end{eqnarray}
which vanishes in the lowest energy limit. The quasiparticle residue
is connected to the real part of retarded self-energy
$\mathrm{Re}\Sigma^{R}(\omega)$ by the relation
\begin{eqnarray}
Z_{f} = \frac{1}{\left|1-\frac{\partial}{\partial\omega}
\mathrm{Re}\Sigma^{R}(\omega)\right|}. \label{Eq:ZfDefinition}
\end{eqnarray}
Using the transformation $\omega = \omega_{0}e^{-\ell}$, where
$\omega_{0}$ is the initial value of energy, we can obtain
\begin{eqnarray}
\mathrm{Re}\Sigma^{R}(\omega)\sim\omega^{1-\frac{5}{23}}. \label{Eq:SelfEnergyTWFPRe}
\end{eqnarray}
Making use of the Kramers-Kronig relation, the imaginary part of
retarded self-energy takes the form
\begin{eqnarray}
\mathrm{Im}\Sigma^{R}(\omega)\sim\omega^{1-\frac{5}{23}}.
\label{Eq:SelfEnergyTWFPIm}
\end{eqnarray}
It is clear that $x$-RVP induces unusual critical behavior at such a
stable fixed point.

\begin{table}[htbp]
\caption{DOS and specific heat of double-WSM. $\gamma_{0}$ represents
zero-energy disorder scattering
rate.\label{Table:ObservableQuantitiesDW}} \vspace{-0.3cm}
\begin{center}
\begin{tabular}{|c|c|c|}
\hline\hline Phase & DOS & Specific heat
\\
\hline Clean & $\rho(\omega)\sim \omega$ & $C_{v}(T)\propto T^{2}$
\\
\hline \, CDM \, & \,
$\rho(0)\sim\gamma_{0}\ln\left(\frac{\Lambda}{\gamma_{0}}\right)$ \, &
$C_{v}(T)\sim \gamma_{0}\ln\left(\frac{\Lambda}{\gamma_{0}}\right)T \sim
\rho(0)T$
\\
\hline\hline
\end{tabular}
\end{center}
\end{table}
\begin{table}[htbp]
\caption{DOS and specific heat of triple-WSM. Here, SCS stands for
stable critical state. \label{Table:ObservableQuantitiesTW}}
\vspace{-0.3cm}
\begin{center}
\begin{tabular}{|c|c|c|}
\hline\hline \quad Phase \quad & \quad DOS \quad & \qquad Specific
heat \qquad
\\
\hline \quad Clean \quad & \quad
$\rho(\omega)\sim\omega^{\frac{2}{3}}$ \qquad & \quad $C_{v}(T)\sim
T^{\frac{5}{3}}$ \qquad
\\
\hline\quad CDM \quad & \quad $\rho(0)\sim \gamma_{0}^{\frac{2}{3}}$
\qquad & \qquad $C_{v}(T)\sim \gamma_{0}^{\frac{2}{3}}T\sim\rho(0) T$
\qquad\qquad
\\
\hline \quad SCS \quad & \quad
$\rho(\omega)\sim\omega^{\frac{7}{23}}$ \qquad  & \qquad
$C_{v}(T)\sim T^{\frac{30}{23}}$ \qquad
\\
\hline\hline
\end{tabular}
\end{center}
\end{table}

The parameters $v$ and $B$ display the low-energy asymptotic
behavior, $v\sim e^{-\frac{5}{23}\ell}$ and $B\sim
e^{-\frac{5}{23}\ell}$, respectively, which in turn leads to
power-law corrections to DOS and specific heat. Formally, the DOS
and specific heat can be written as
\begin{eqnarray}
\rho(\omega)&\sim&\omega^{\frac{7}{23}}, \label{Eq:DOSTWFP}
\\
C_{v}(T)&\sim& T^{\frac{30}{23}}. \label{Eq:CvTWFP}
\end{eqnarray}

Here, we would like to remark on the interpretation of the above
result. The terminology of FL or NFL is widely used in the
literature to describe an interacting fermion system. If the
fermions are subject to an inelastic interaction, which could be
caused by the Coulomb potential, gauge fields, or the quantum
fluctuation of certain order parameter \cite{Varma02, Lohneysen,
LeePA06, LeeSS09, Abrikosov74, Moon13, Herbut14, Janssen, Isobe16},
the resultant Landau damping rate $\gamma(\omega)\propto
\left|\mathrm{Im}\Sigma^{R}(\omega)\right|$ must vanish as $\omega
\rightarrow 0$, required by the Pauli's exclusion rule. The
influence of static disorder relies sensitively on the running
property of the disorder strength parameter. In ordinary metals and
certain SM materials, including triple-WSM, RSP is usually a
relevant perturbation and thus generates a disorder scattering rate
$\gamma(\omega)$, which approaches a finite constant $\gamma_0$ at
$\omega = 0$. The fermion system with a nonzero $\gamma_0$ is
normally identified as a diffusive metal. In distinction to RSP,
$x$-RVP is a marginal perturbation in triple-WSM, and it induces a
stable infrared fixed point. At this fixed point, the fermion field
operator acquires a finite anomalous dimension, the residue $Z_f$
vanishes as $\omega \rightarrow 0$, and both DOS and specific heat
display power-law behavior. All these characteristics resemble those
of a typical NFL induced by inelastic interactions. We notice that
such kind of disorder-induced state might be realized in a variety
of SM materials. For instance, it could be produced in 2D Dirac
fermion systems by RVP \cite{Ludwig94, Nersesyan94, Altland02,
Ostrovsky06, Foster12, Foster14, JingWang17} or random mass with
long-range correlation \cite{Fedorenko12}. A 3D DSM/WSM that is
close to SM-CDM QCP may be driven by RSP to flow to such a state,
which was recently identified as a NFL state \cite{Fu17, Roy16B,
Roy16C, Pixley16A}.

\begin{figure}[h]
\center
\includegraphics[width=3.35in]{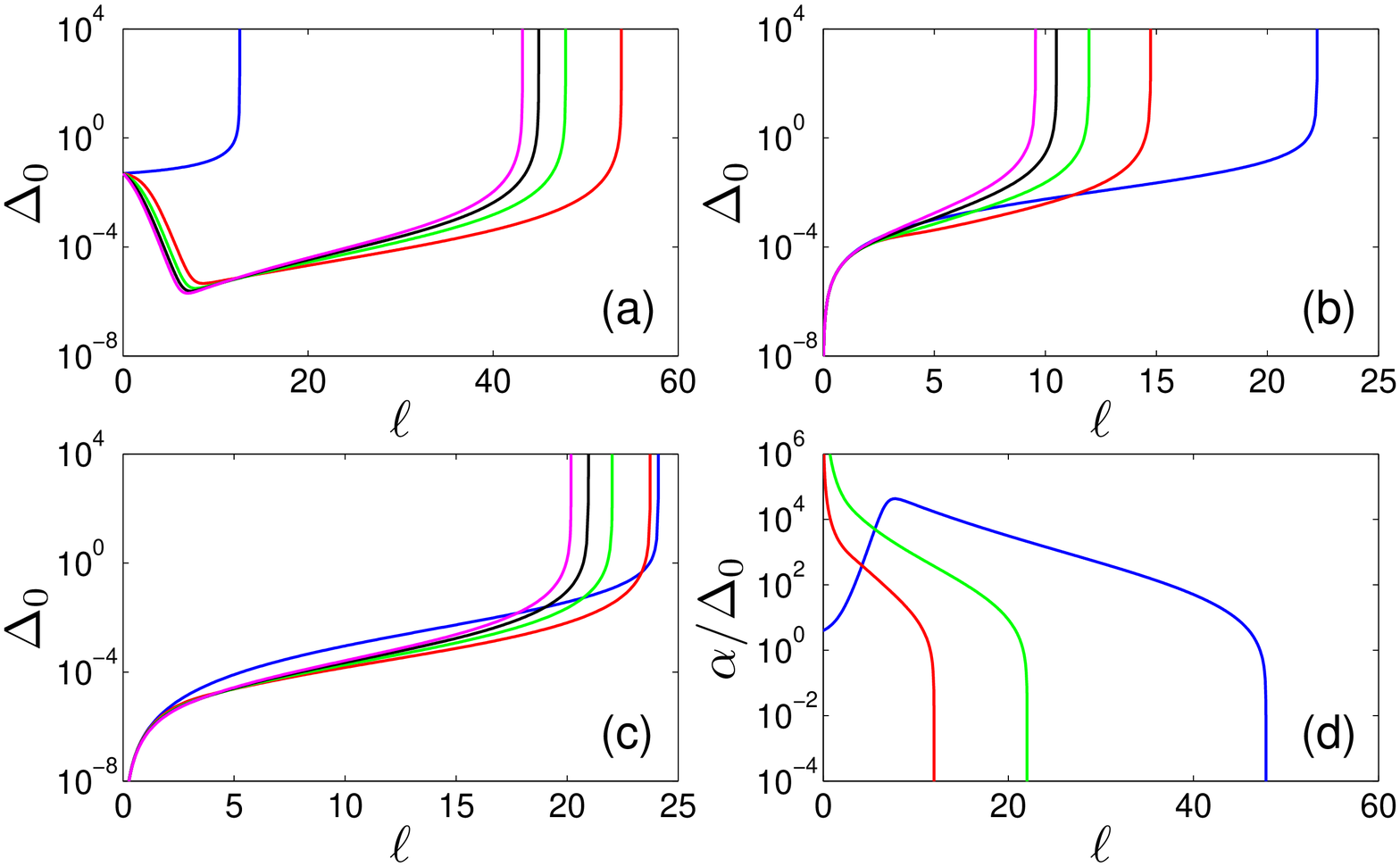}
\caption{Flow of $\Delta_{0}$ caused by the interplay of RSP and
Coulomb interaction in double-WSM, where $\Delta_{00}=0.05$ in (a),
$\Delta_{10}=0.05$ in (b), and $\Delta_{30}=0.05$ in (c). In
(a)-(c), the blue, red, green, black, and magenta curves represent
the initial values $\alpha_{0} = 0, 0.1, 0.2, 0.3, 0.4$,
respectively. Dependence of the ratio $\alpha_{0}/\Delta_{0}$ on
$\ell$ is shown in (d) with $\alpha_{0} = 0.2$, where the blue, red,
and green curves represent the values of $\Delta_{00}=0.05$,
$\Delta_{10}=0.05$, $\Delta_{30}=0.05$, respectively. An initial
value $\zeta_{0}^{d}=0.1$ is taken.\label{Fig:RGDW}}
\end{figure}

The stable fixed point of triple-WSM induced by the $x$-RVP exhibit
similar behavior to the NFLs. In a loose sense, this state could be
called a NFL \cite{Fu17, Roy16B, Roy16C, Pixley16A}. However, we
should remember that, strictly speaking, the coupling between
fermions and static disorder $x$-RVP is rather different from
inelastic scattering. To reflect this difference, it might be more
appropriate to use a different terminology. Here, we would consider
the NFL-like behavior characterized by Eq.~(\ref{Eq:ZfTWFP}) and
Eqs.~(\ref{Eq:SelfEnergyTWFPRe})-(\ref{Eq:CvTWFP}) as stable
critical behavior.

When $y$-RVP exists alone in the system, there is also a stable
infrared fixed point, at which the quantum critical behavior is
analogous to that caused by $x$-RVP. Thus, we will not further
discuss the physical effects of $y$-RVP.

If the system contains only $z$-RVP, the RG equation for its
strength parameter is given by
\begin{eqnarray}
\frac{d\Delta_{3}}{d\ell} = \frac{1}{3}\Delta_{3} +
\frac{1}{12}\Delta_{3}^{2}.
\end{eqnarray}
The corresponding solution has the form
\begin{eqnarray}
\Delta_{3} = \frac{4e^{\frac{1}{3}\ell}\Delta_{30}}{\Delta_{30} + 4 -
e^{\frac{1}{3}\ell}\Delta_{30}}.
\end{eqnarray}
It is clear that $\Delta_{3}$ flows to infinity at some finite
energy scale if $\Delta_{30}$ takes any finite value. Therefore, the
$z$-RVP is similar to RSP, and an arbitrarily weak $z$-RVP is able
to drive a transition to CDM phase.

The above results inform us that the disorder effects on double- and
triple-WSMs are very different: disorder always drives a CDM state
of double-WSM, but can induce either CDM state or stable critical
state, depending on the specific type of disorder. To better
understand the difference, we list the asymptotic behavior of DOS
and specific heat in different phases of double- and triple-WSMs in
Tables~\ref{Table:ObservableQuantitiesDW} and
\ref{Table:ObservableQuantitiesTW}, respectively.

\begin{figure}[htbp]
\center
\includegraphics[width=3.35in]{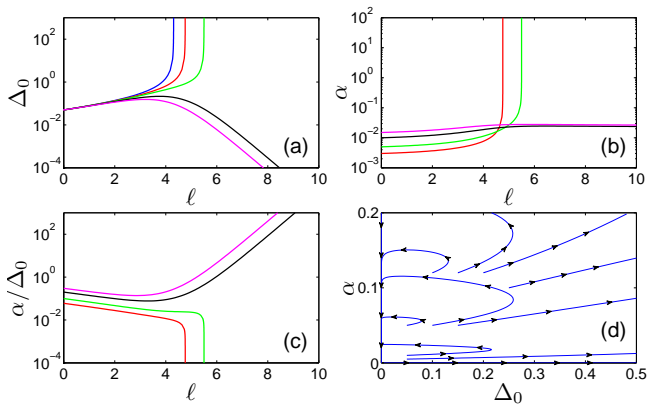}
\caption{(a)-(c): Flow of $\Delta_{0}$, $\alpha$, and
$\alpha/\Delta_{0}$ due to interplay of RSP and Coulomb interaction
in triple-WSM at a given value $\Delta_{00}=0.05$. Blue, red, green,
black, and magenta curves represent the initial values
$\alpha_{0}=0, 0.003, 0.005, 0.01, 0.015$, respectively. (d) shows
the flowing diagram on the $\Delta_{0}$-$\alpha$ plane. Here, we
assume that $\zeta_{0}^{t}=0.1$. \label{Fig:RGTWIni0}}
\end{figure}

\subsection{Interplay between interaction and disorder}

We now consider the influence of the interplay between Coulomb
interaction and disorder.

For double-WSM, we present the flow of $\Delta_{0}$ with
$\Delta_{00}=0.05$ and $\Delta_{10}=\Delta_{20}=\Delta_{30}=0$ in
Fig.~\ref{Fig:RGDW}(a). In the presence of Coulomb interaction,
$\Delta_{0}$ first decreases with increasing $\ell$, but then starts
to increase once $\ell$ exceeds certain threshold. As $\ell$
continues to grow, $\Delta_{0}$ finally flows to strong coupling
regime at some finite energy scale. In Fig.~\ref{Fig:RGDW}(b) and
Fig.~\ref{Fig:RGDW}(c), we show the curves of $\Delta_{0}$ obtained
by assuming that the system contains, apart from Coulomb
interaction, only $x$-RVP and only $z$-RVP, with initial values
$\Delta_{10}=0.05$ and $\Delta_{30}=0.05$, respectively. An
interesting result is that $\Delta_{0}$ is always dynamically
generated and flows to strong coupling at sufficiently large $\ell$.

According to Figs.~\ref{Fig:RGDW}(a)-(c), irrespective of the initial
value of Coulomb interaction strength, $\Delta_{0}$ always flows to
strong coupling if any type of disorder has a finite strength. It is
found that $\alpha$ also goes to the strong coupling regime, which
is not depicted in Fig.~\ref{Fig:RGDW}, but the ratio
$\alpha/\Delta_{0} \rightarrow 0$ in any case, as clearly shown in
Fig.~\ref{Fig:RGDW}(d). This indicates that disorder is always more
important than Coulomb interaction, and the low-energy properties of
the system are mainly determined by the disorder, rather than by the
Coulomb interaction. It was also found that RSP dominates over any
component of RVP. An immediate conclusion is that double-WSM is
always in the CDM phase, no matter the Coulomb interaction is
incorporated or not. This is similar to the case of the 3D
quadratical SM, which is found by Nandkishore \emph{et al.}
\cite{Nandkishore17} to be always in the CDM phase when Coulomb
interaction and disorder are simultaneously present.

\begin{figure}[htbp]
\center
\includegraphics[width=3.35in]{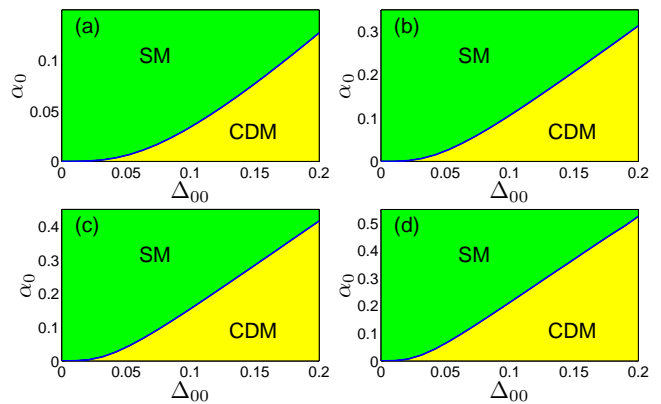}
\caption{Phase diagrams of the triple-WSM in the presence of RSP and
Coulomb interaction, plotted on the plane spanned by $\Delta_{00}$
and $\alpha_0$. The value of $\zeta_{0}^{t}$ is taken to be $0.1$,
$0.5$, $1$, and $2$ in (a), (b), (c), and (d), respectively. There
is always a critical line that separates the SM and CDM phases.
\label{Fig:PhaseDiagramRSPTW}}
\end{figure}

For triple-WSM, we show in Figs.~\ref{Fig:RGTWIni0}(a)-(c) the flows
of parameters $\Delta_{0}$, $\alpha$, and $\alpha/\Delta_{0}$
obtained by considering both RSP and Coulomb interaction.
Fig.~\ref{Fig:RGTWIni0}(d) presents the flow diagram of the two
parameters $\Delta_{0}$ and $\alpha$. We find that, for weak Coulomb
interaction, $\Delta_{0}$ and $\alpha$ still flow to strong
couplings at some finite scale, but the ratio $\alpha/\Delta_{0}
\rightarrow 0$, implying the dominance of RSP at low energies.
However, if $\alpha_{0}$ is greater than a critical value, whose
precise value is determined by $\Delta_{00}$, the parameters
$\Delta_{0}$, $\alpha$, and also $\Delta_{0}/\alpha$ flow to zero at
lowest energy limit. Apparently, the triple-WSM recovers a SM phase
with vanishing $\rho(0)$ when the initial value of Coulomb
interaction becomes sufficiently strong. In short, RSP is much more
important than the weak Coulomb interaction, driving the system to
become CDM, but the strong Coulomb interaction plays an overwhelming
role than RSP and guarantees the stability of the SM
phase. We notice that such behavior is very similar to that of 2D
DSM caused by the interplay between the RSP and Coulomb interaction
\cite{Ye99, Stauber05, WangLiu14}.

The phase diagram of triple-WSM in the plane of $\Delta_{00}$ and
$\alpha_{0}$  is shown in Fig.~\ref{Fig:PhaseDiagramRSPTW}.
Obviously, there is a critical line, at which a QPT takes place
between the SM and CDM phases. In the
Figs.~\ref{Fig:PhaseDiagramRSPTW}(a)-(d), $\zeta_{0}$ is taken
$0.1$, $0.5$, $1$, and $2$ respectively. We can find that the change
of $\zeta_{0}$ does not change the qualitative characteristic of the
phase diagram, but quantitatively modify the critical line between
SM and CDM phase. For ordinary WSMs, Goswami \emph{et al.}
\cite{Goswami11} showed that there is also an analogous critical
line between the SM and CDM phases in the plane spanned by the
initial strength parameters of Coulomb interaction and RSP. However,
the crossover point of the critical line and axes is $(0, 0)$ for
triple-WSM, but $(\Delta_{00}^{c}, 0)$ for ordinary WSM, where
$\Delta_{00}^{c}$ is a finite value. This feature arises from the
fact that an arbitrarily weak RSP drives the triple-WSM to enter
into a CDM phase if there is only RSP. The situation is different in
the usual WSM, where only a sufficiently strong RSP can drive a CDM
transition \cite{Goswami11}.

\begin{figure}[htbp]
\center
\includegraphics[width=3.35in]{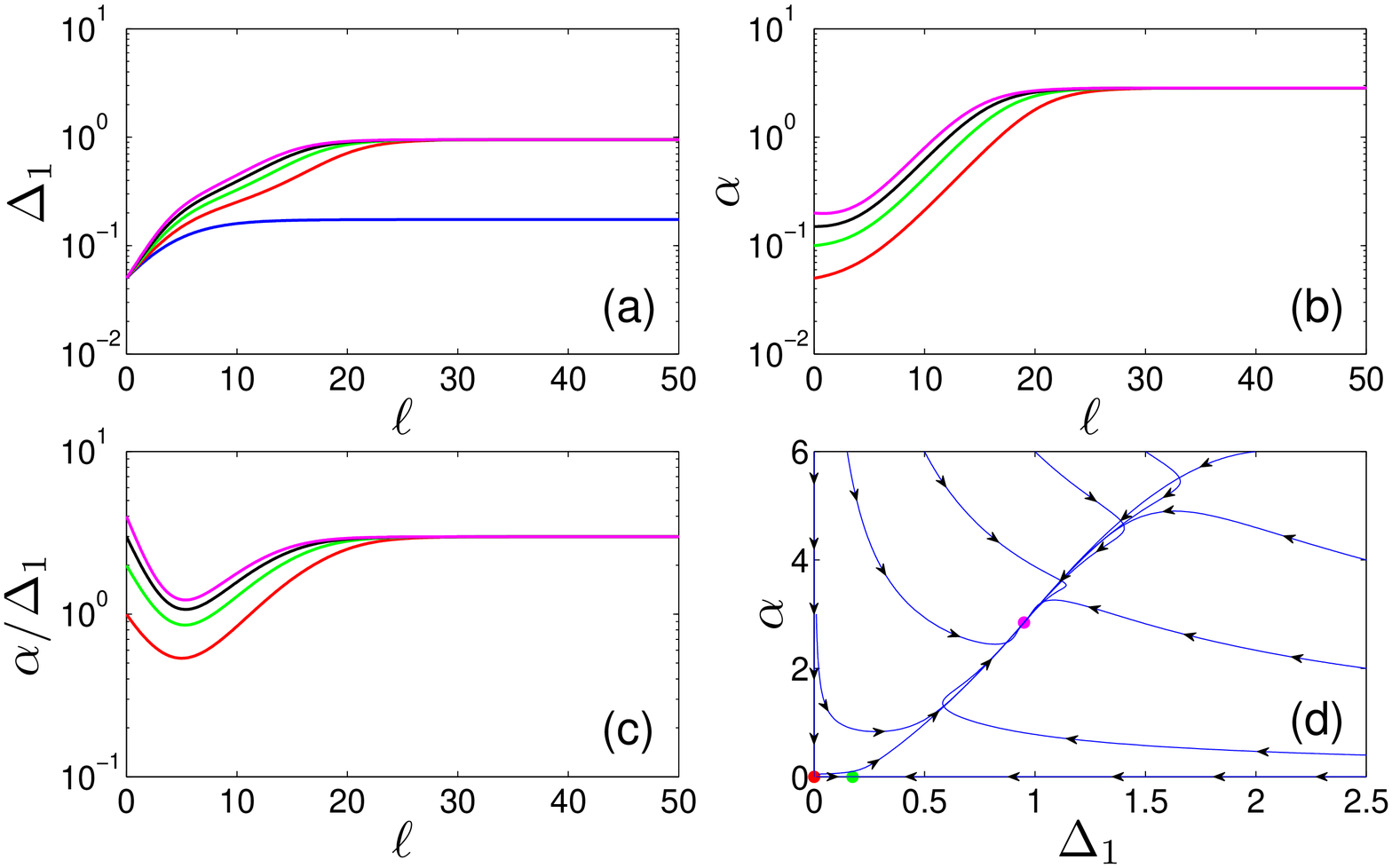}
\caption{(a)-(c): Flow of $\Delta_{1}$, $\alpha$, and
$\alpha/\Delta_{1}$ caused by the interplay of $x$-RVP and Coulomb
interaction in triple-WSM at a given value $\Delta_{10} = 0.05$.
Blue, red, green, black, and magenta curves correspond to
$\alpha_{0}=0, 0.05, 0.1, 0.15, 0.2$, respectively. Flowing diagram
on the $\Delta_{1}$-$\alpha$ plane is shown in (d). Here, we choose
an initial value $\zeta_{0}^{t}=0.1$. \label{Fig:RGTWIni1}}
\end{figure}

We then consider the mutual influence of $x$-RVP and Coulomb
interaction. According to Fig.~\ref{Fig:RGTWIni1}, $\Delta_{1}$ and
$\alpha$ always flow to a stable infrared fixed point
(${\Delta_{1n}^{*}},\alpha_{n}^{*}$). This stable fixed point is
possibly characterized by the emergence of unusual critical
behavior, which is physically distinct from both SM and CDM states.
We emphasize that the existence and property of such a fixed point
needs to be further explored, because $\Delta_{1n}^{*}$ is only
slightly smaller than unity whereas $\alpha_{n}^{*} \approx 3$. At
such a fixed point, the validity of perturbative RG calculations is
actually questionable. We expect that other non-pertubative method,
such as functional RG (fRG) \cite{Metzner12, Bauer15, Sharma16,
SbierskifRG17} or Monte Carlo simulation \cite{Foulkes01, Gull11,
Tupitsyn17}, could be employed to further study this problem. When
there is an interplay of Coulomb interaction and $y$-RVP, the
low-energy properties are very similar to the case of $x$-RVP, and
thus will not be further discussed.

\begin{figure}[htbp]
\center
\includegraphics[width=3.35in]{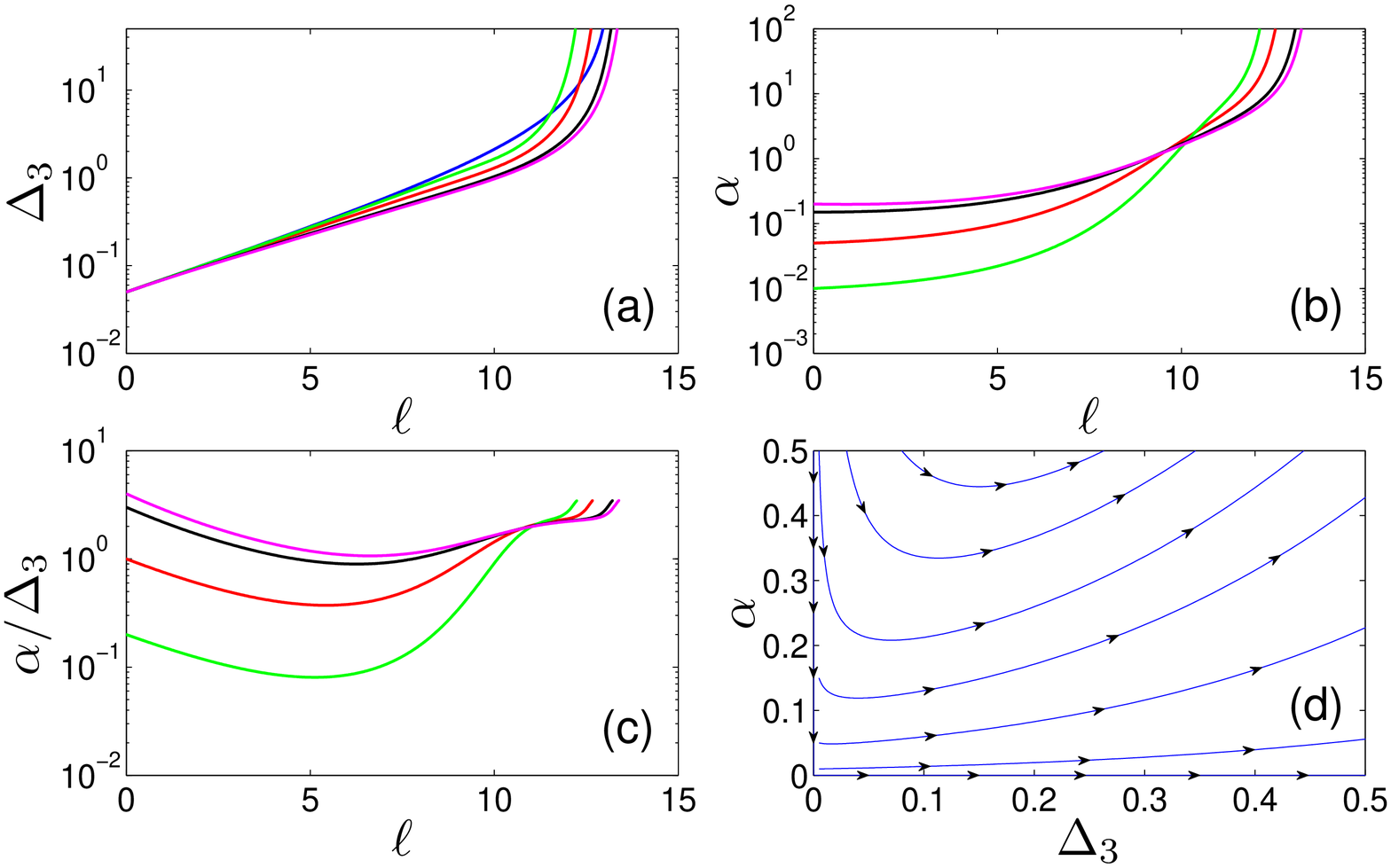}
\caption{(a)-(c): Flow of $\Delta_{3}$, $\alpha$, and
$\alpha/\Delta_{3}$ caused by the interplay of $z$-RVP and Coulomb
interaction in triple-WSM. In (a)-(c), $\Delta_{30}=0.05$, and the
blue, red, green, black, and magenta curves represent the initial
values $\alpha_{0}=0, 0.05, 0.1, 0.15, 0.2$, respectively. (d) shows
the flowing diagram in the $\Delta_{3}$-$\alpha$ plane. An initial
value $\zeta_{0}^{t}=0.1$ is taken. \label{Fig:RGTWIni3}}
\end{figure}

We finally consider the interplay of Coulomb interaction and
$z$-RVP. The RG solutions for parameters $\Delta_{3}$, $\alpha$, and
$\alpha/\Delta_{3}$ are depicted in Figs.~\ref{Fig:RGTWIni3}(a)-(c).
The schematic flowing diagram in the $\Delta_{3}-\alpha$ plane is
plotted in Fig.~\ref{Fig:RGTWIni3}(d). According to these results,
we find that $\Delta_{3}$ and $\alpha$ both flow to the strong
coupling regime, where $\alpha$ dominates over $\Delta_{3}$.
Therefore, the Coulomb interaction plays a more important role than
$z$-RVP at low energies. A possible interpretation of such behavior
is that the system becomes an interaction-dominated Mott insulator.
This is an important issue that deserves further theoretical
investigation.

\begin{figure}[htbp]
\center
\includegraphics[width=3.35in]{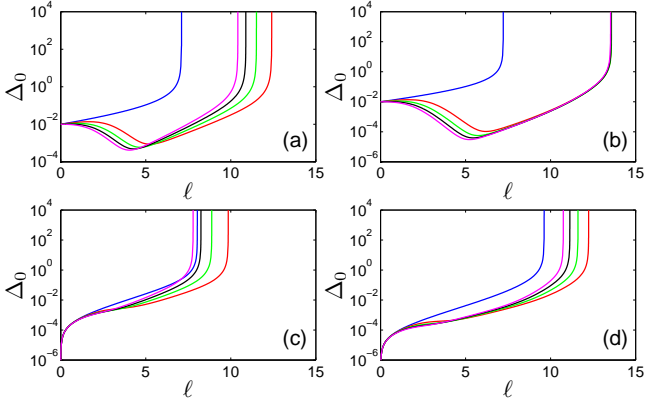}
\caption{(a)-(c): Flow of $\Delta_{0}$ driven by two types of
disorder and Coulomb interaction in triple-WSM. (a):
$\Delta_{00}=0.01$, $\Delta_{10}=0.01$; (b): $\Delta_{00}=0.01$,
$\Delta_{30}=0.01$; (c): $\Delta_{10}=0.01$, $\Delta_{20}=0.01$;
(d): $\Delta_{10}=0.01$, $\Delta_{30}=0.01$. Blue, red, green,
black, and magenta curves represent the initial values
$\alpha_{0}=0, 0.05, 0.1, 0.15, 0.2$, respectively. Here, we set
$\zeta_{0}^{t} = 0.1$. \label{Fig:RGMixtureTW}}
\end{figure}

\begin{figure}[htbp]
\center
\includegraphics[width=3.35in]{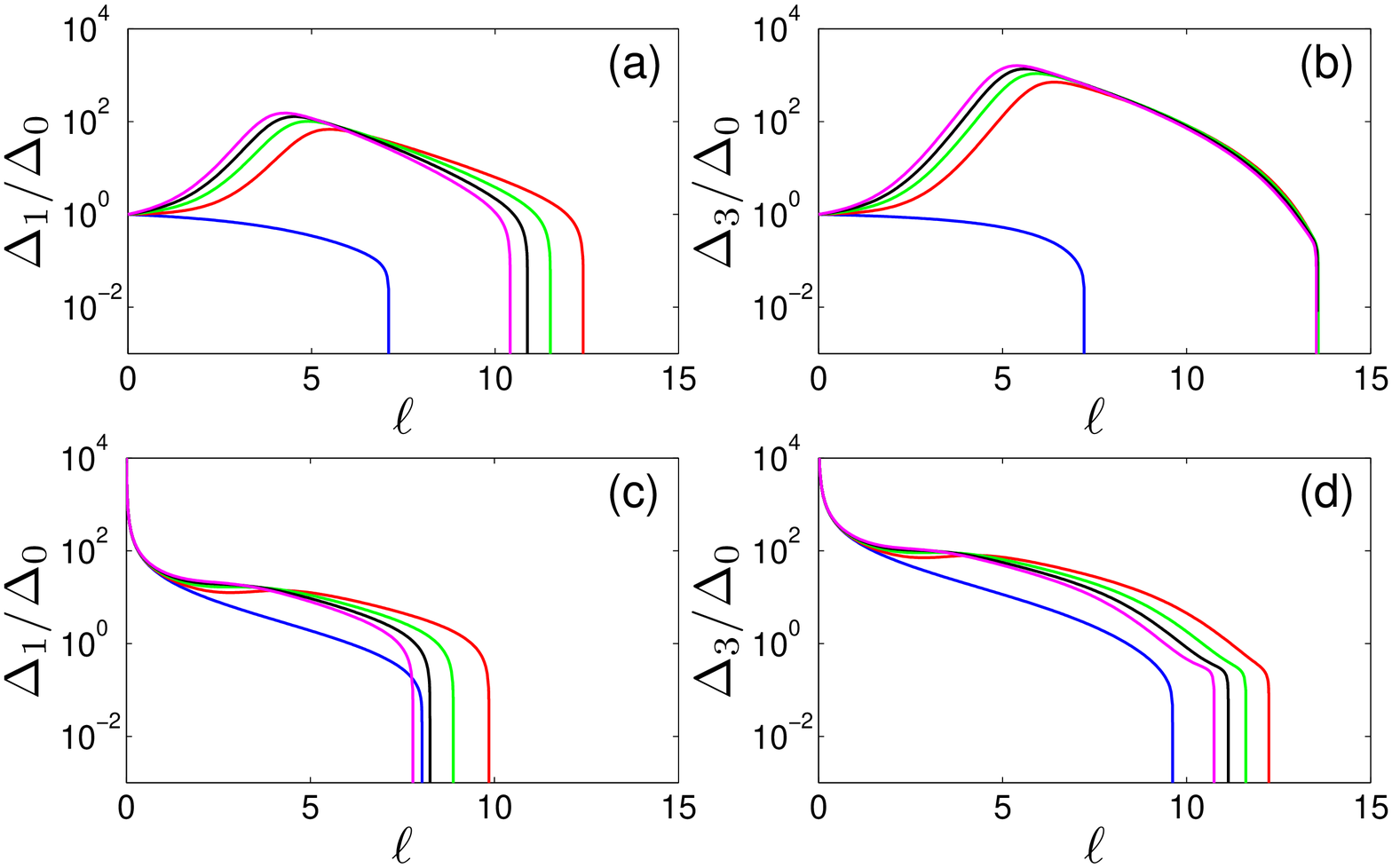}
\caption{(a)-(d): Flow of $\Delta_{1}/\Delta_{0}$ in (a), (c), and
$\Delta_{3}/\Delta_{0}$ in (b) and (d). The initial conditions are
the same as Fig.~\ref{Fig:RGMixtureTW}.\label{Fig:RGMixtureRatioTW}}
\end{figure}

For triple-WSM, if two types of disorder are initially considered,
other types of disorder are always generated, and all the disorder
parameters flow to infinity at finite energy scale. The running
behavior of $\Delta_{0}$ by considering the coexistence of two types
of disorder is presented in Fig.~\ref{Fig:RGMixtureTW}, which
clearly shows that $\Delta_{0}$ always flows to the strong coupling
regime. The strength of Coulomb interaction $\alpha$ also flows to
infinity at the same energy scale. However, the ratios
$\Delta_{1}/\Delta_{0}$, $\Delta_{2}/\Delta_{0}$,
$\Delta_{3}/\Delta_{0}$ and $\alpha/\Delta_{0}$ all decrease down to
zero. From Fig.~\ref{Fig:RGMixtureRatioTW}, obtained under the same
initial conditions as Fig.~\ref{Fig:RGMixtureTW}, we observe that
both $\Delta_{1}/\Delta_{0}$ or $\Delta_{3}/\Delta_{0}$ vanish at
finite $\ell$. Therefore, the low-energy physics is dominated by
RSP, and the system is inevitably turned into the CDM phase. This
implies that triple-WSM always becomes a CDM if two or more types of
disorder exist simultaneously, which is similar to double-WSM. It is
interesting that similar phenomena occur in 2D DSM \cite{Evers08,
Ostrovsky06, Foster12, JingWang17}, where RSP, RVP, and random mass
can induce CDM transition, stable critical state, and
logarithmic-like corrections to observable quantities, respectively.
If any two types of disorder coexist in 2D DSM, the system always
undergoes a CDM transition.

\section{Further analysis of RG results\label{Sec:Discussions}}

In this section, we present a further analysis of the RG results
obtained and discussed in the last section.

\subsection{Difference between double- and triple-WSMs}

An important indication of the analysis presented in
Section~\ref{Sec:Results} is that double- and triple-Weyl fermions
manifest very different low-energy behavior in response to static
short-range disorder. In the following, we explain the physical
origin of such marked difference, and also compare with the usual
WSM.


The free propagators for the usual, double-, and triple-Weyl
fermions are given by
\begin{eqnarray}
G_{0}^{u}(\omega,\mathbf{k}) &=& \frac{1}{-i\omega +
v\left(k_{x}\sigma_{1}+k_{y}\sigma_{2}+k_{z}\sigma_{3}\right)},
\nonumber \\
G_{0}^{d}(\omega,\mathbf{k})&=&\frac{1}{-i\omega+Ad_{1}(\mathbf{k})
\sigma_{1} + Ad_{2}(\mathbf{k})\sigma_{2}+vk_{z}\sigma_{3}},
\nonumber \\
G_{0}^{t}(\omega,\mathbf{k}) &=& \frac{1}{-i\omega +
Bg_{1}(\mathbf{k})\sigma_{1} + Bg_{2}(\mathbf{k})\sigma_{2} +
vk_{z}\sigma_{3}}, \nonumber
\end{eqnarray}
where $d_{1}(\mathbf{k})=k_{x}^{2}-k_{y}^{2}$,
$d_{2}(\mathbf{k})=2k_{x}k_{y}$,
$g_{1}(\mathbf{k})=k_{x}^{3}-3k_{x}k_{y}^{2}$, and
$g_{2}(\mathbf{k})=k_{y}^{3}-3k_{y}k_{x}^{2}$. For usual WSM, the
fermion propagator satisfies the relation
\begin{eqnarray}
G_{0}^{u}(\omega,\mathbf{k}) + G_{0}^{u}(-\omega,-\mathbf{k}) = 0.
\label{Eq:WeylPropagatorCondition}
\end{eqnarray}
It is easy to verify that the fermion propagator in double-WSM does
not satisfy this relation, i.e.,
\begin{eqnarray}
G_{0}^{d}(\omega,\mathbf{k}) + G_{0}^{d}(-\omega,-\mathbf{k})\neq 0.
\label{Eq:DoubleWeylPropagatorCondition}
\end{eqnarray}
For triple-WSM, one finds that
\begin{eqnarray}
G_{0}^{t}(\omega,\mathbf{k}) + G_{0}^{t}(-\omega,-\mathbf{k}) = 0,
\label{Eq:TripleWeylPropagatorCondition}
\end{eqnarray}
which shares the same property as usual WSM.

The diagrams for one-loop corrections to fermion-disorder vertex are
given by Fig.~\ref{Fig:VertexCorrection} in the
Appendix~\ref{App:DisVertexCorrection}. Amongst these diagrams, the
sum of (b) and (c) are
\begin{eqnarray}
V_{ij}^{u,d,t(2)+(3)} &=& 2\Delta_{i}\Delta_{j}
\int'\frac{d^3\mathbf{k}}{(2\pi)^{3}}\left(\psi_{a}^{\dag}\Gamma_{i}
G_{0}^{u,d,t}(0,\mathbf{k})
\Gamma_{j}\psi_{a}\right)\nonumber \\
&&\times\left\{\psi_{b}^{\dag}
\left[\Gamma_{j}G_{0}^{u,d,t}(0,\mathbf{k})
\Gamma_{i}\right.\right.\nonumber
\\
&&\left.\left. + \Gamma_{i}G_{0}^{u,d,t}(0,-\mathbf{k})
\Gamma_{j}\right]\psi_{b}\right\}.\label{Eq:VertexCorrectionB2}
\end{eqnarray}
In the case of usual WSM with $\Gamma_{i} = \Gamma_{j}$, the
constraint of Eq.~(\ref{Eq:WeylPropagatorCondition}) ensures that
the total contribution from these two Feynman diagrams vanishes,
namely
\begin{eqnarray}
V_{ii}^{u(2)+(3)} = 0. \label{Eq:WeylTwoDiagramContribution}
\end{eqnarray}
For double-WSM with $\Gamma_{i}=\Gamma_{j}$, we know from
Eq.~(\ref{Eq:DoubleWeylPropagatorCondition}) that the total
contribution
\begin{eqnarray}
V_{ii}^{d(2)+(3)}\neq 0, \label{Eq:DoubleWeylTwoDiagramContribution}
\end{eqnarray}
which differs from usual WSM. More concretely, we have
\begin{eqnarray}
V_{00}^{(2)+(3)} &=& \frac{\Delta_{0}^{2}}{8\pi
vA}\sum_{j=1,2}\left(\psi_{a}^{\dag}\sigma_{j} \psi_{a}\right)
\left(\psi_{b}^{\dag}\sigma_{j}\psi_{b}\right)\ell,
\label{Eq:DoubleWeylTwoDiagramContributionA}
\\
V_{11}^{(2)+(3)} &=& \frac{\Delta_{1}^{2}}{8\pi
vA}\sum_{j=1,2}\left(\psi_{a}^{\dag}\sigma_{j}\psi_{a}\right)
\left(\psi_{b}^{\dag}\sigma_{j}\psi_{b}\right)\ell,
\label{Eq:DoubleWeylTwoDiagramContributionB}
\\
V_{22}^{(2)+(3)} &=& \frac{\Delta_{2}^{2}}{8\pi
vA}\sum_{j=1,2}\left(\psi_{a}^{\dag}\sigma_{j}\psi_{a}\right)
\left(\psi_{b}^{\dag}\sigma_{j}\psi_{b}\right)\ell,
\label{Eq:DoubleWeylTwoDiagramContributionC}
\\
V_{33}^{(2)+(3)} &=& \frac{\Delta_{3}^{2}}{8\pi
vA}\sum_{j=1,2}\left(\psi_{a}^{\dag}\sigma_{j}\psi_{a}\right)
\left(\psi_{b}^{\dag}\sigma_{j}\psi_{b}\right)\ell.
\label{Eq:DoubleWeylTwoDiagramContributionD}
\end{eqnarray}
The triple-WSM is very similar to the usual WSM, which means that,
in the case $\Gamma_{i} = \Gamma_{j}$, the total contribution of
Figs.~\ref{Fig:VertexCorrection}(b) and (c) satisfies
\begin{eqnarray}
V_{ii}^{t(2)+(3)} = 0,\label{Eq:TripleWeylTwoDiagramContribution}
\end{eqnarray}
as a direct results of Eq.~(\ref{Eq:TripleWeylPropagatorCondition}).

The close analogy between usual and triple-WSMs, reflected in the
constraints given by Eqs.~(\ref{Eq:WeylTwoDiagramContribution}) and
(\ref{Eq:TripleWeylTwoDiagramContribution}), leads to common
properties shared by these two systems. For instance, one type of
disorder can exist individually in usual and triple-WSMs. However,
this is not possible in a double-WSM. Due to the non-zero
contributions shown in
Eqs.~(\ref{Eq:DoubleWeylTwoDiagramContributionA})-(\ref{Eq:DoubleWeylTwoDiagramContributionD}),
one type of disorder cannot exist individually because it
dynamically generates other types of disorder \cite{BeraRoy16}. We
have pointed out such special property of double-WSM in
Section~\ref{Sec:Results}. From the above analysis, we can now
conclude that the specific disorder scattering processes represented
by Figs.~\ref{Fig:VertexCorrection}(b) and (c) are suppressed in
both the usual and triple-WSMs, where the Hamiltonian
$\mathcal{H}(\mathbf{k})$ becomes $-\mathcal{H}(\mathbf{k})$ under
the transformation $\mathbf{k}\rightarrow -\mathbf{k}$, but make
non-trivial contributions in the case of double-WSM.

While the disorder effects on the usual and triple-WSMs are similar
in some aspects, they are definitely not identical. In a usual WSM
containing only one component of RVP, Sbierski \emph{et al.}
\cite{Sbierski16} showed that the disorder strength parameter
satisfies the equation
\begin{eqnarray}
\frac{d\Delta_{i}}{d\ell} = -\Delta_{i}-C\Delta_{i}^{2},
\label{Eq:RGRVWeyl}
\end{eqnarray}
where $i=1$, $2$, or $3$, and $C$ is a positive constant. For the
triple-WSM with only $x$- or $y$-component of RVP, we have showed in
Sec.~\ref{Sec:Results} that the RG equation is
\begin{eqnarray}
\frac{d\Delta_{i}}{d\ell} = \frac{1}{3}\Delta_{i} -
\frac{23}{12}\Delta_{i}^{2},\label{Eq:RGRVTripleWeyl}
\end{eqnarray}
where $i=1$ or $2$. The second terms of the right hand sides of
Eq.~(\ref{Eq:RGRVWeyl}) and Eq.~(\ref{Eq:RGRVTripleWeyl}),
representing the one-loop correction to beta function, are both
negative, which is valid if the system contains only one component
of RVP and the relations given by
Eq.~(\ref{Eq:WeylPropagatorCondition}) and
Eq.~(\ref{Eq:TripleWeylPropagatorCondition}) are satisfied. However,
the first terms, determined by the scaling dimension at tree-level,
are opposite in sign. This difference is owing to the fact that the
usual Weyl and triple-Weyl fermions have different dispersions.
According to Eq.~(\ref{Eq:RGRVWeyl}), we know that the disorder
strength for one component of RVP always flows to zero at low
energies in the usual WSM \cite{Sbierski16}. If triple-WSM
containing only $x$- or $y$-component of RVP, there is a stable
fixed point $\Delta_{i}^{*} = \frac{4}{23}$, which is obtained from
Eq.~(\ref{Eq:RGRVTripleWeyl}).

For double-WSM, the situation is in sharp contrast to the usual and
triple-WSMs. As demonstrated in the last section, the double-WSM is
always driven by $x$-RVP (or $y$-RVP) to enter into a CDM phase.
This result should be attributed to
Eq.~(\ref{Eq:DoubleWeylPropagatorCondition}).

\subsection{Stability of the infrared fixed point of triple-WSM}

In Section~\ref{Sec:Results}, we have found a stable infrared fixed
point in our one-loop RG analysis of the triple-WSM containing only
the $x$- or $y$-component of RVP. A natural question arises as
whether such a fixed point survives the higher order corrections. To
address this issue, we now discuss whether the one-loop results are
still valid after including two-loop corrections.

It is useful to first briefly review the recent progress of disorder
effects in 3D DSM/WSM. For 3D DSM and WSM, one-loop RG studies
\cite{Goswami11, Roy14, Syzranov15A, Syzranov15B} revealed that weak
RSP is irrelevant, but becomes relevant if the RSP strength is
beyond a critical value, which then drives a SM-CDM transition. To
the one-loop order, the dynamical critical exponent at quantum
critical point (QCP) from SM to CDM is $z = \frac{3}{2}$, and the
correlation length exponent is $\nu = 1$ \cite{Goswami11, Roy14,
Syzranov15A, Syzranov15B}. Two-loop order corrections were also
calculated by various approaches, including replica method
\cite{Roy14, Roy16Erratum}, supersymmetry technique
\cite{Syzranov16A}, correspondence with Gross-Neveu model
\cite{Roy16Erratum}, and correspondence with Gross-Neveu-Yukawa
model \cite{Louvet16}. These studies confirmed that there is still a
quantum phase transition (QPT) from SM to CDM, which implies that
the conclusion obtained by one-loop RG analysis is qualitatively
robust against higher order corrections. Quantitatively, the
critical disorder strength, dynamical critical exponent $z$, and
correlation length exponent $\nu$ are more or less modified after
including two-loop corrections. Syzranov \emph{et al.}
\cite{Syzranov16A} found $z \approx 1.4$ and $\nu\approx 0.67$ after
making two-loop calculations. Roy and Das Sarma \cite{Roy16Erratum}
also reported that $\nu \approx 0.67$ up to two-loop order. Louvet
\emph{et al.} \cite{Louvet16} got a nearly identical value $\nu
\approx 0.65-0.67$ in their two-loop calculations. The same problem
has also been investigated by using the numerical simulation method
\cite{BeraRoy16, Kobayashi14, Liu16, Pixley16A, Roy16B, Sbierski14,
Sbierski15, Fu17}, where it is found that a SM-CDM transition always
occurs, providing further support to the conclusion reached by the
one-loop RG analysis. Moreover, some numerical studies
\cite{Pixley16B, Pixley16C, Pixley17} suggested that the rare region
effect can induce exponentially small zero-energy DOS $\rho(0)$ in
the case of weak disorder, which broadens the QCP to a quantum
critical region at finite energy-scale. Actually, the dynamical
critical exponent $z$ obtained by most of the existing numerical
studies \cite{BeraRoy16, Kobayashi14, Sbierski15, Liu16, Pixley16A,
Roy16B, Fu17} is well consistent with the one-loop RG result $z =
\frac{3}{2}$. Although the precise value of $\nu$ is still
controversial \cite{BeraRoy16, Kobayashi14, Liu16, Pixley16A,
Roy16B, Sbierski15, Fu17}, the existing extensive analytical and
numerical works suggest that one-loop RG results are at least
qualitatively reliable.

In a recent work, Sbierski \emph{et al.} \cite{Sbierski16} studied
the impact of RVP on 3D WSM by making a one-loop RG analysis along
with numerical simulation. Their one-loop RG result is that 3D WSM
is always in the SM phase if the system contains only one component
of RVP, and their numerical simulations found that 3D WSM stays in
the SM phase even if RVP becomes very strong, which is well
consistent with the one-loop RG result.

In the case of triple-WSM, the strong anisotropy in the fermion
dispersion makes it very difficult to carry out an analytical
calculation of two-loop corrections to the RG equations. We would
like to leave this for future study. To estimate the possible impact
of two-loop contributions on the one-loop conclusion, we now present
a generic analysis. After including two-loop corrections, the RG
equation could be formally written as
\begin{eqnarray}
\frac{d\Delta_{i}}{d\ell} = \frac{1}{3}\Delta_{i} -
\frac{23}{12}\Delta_{i}^{2} + C_{2\mathrm{LP}}\Delta_{i}^{3},
\label{Eq:RGRVTripleWeylTwoLoop}
\end{eqnarray}
where $i=1$ or $i=2$. The first and second terms on the right hand
side of Eq.~(\ref{Eq:RGRVTripleWeylTwoLoop}) represent the
tree-level and one-loop contributions, respectively, whereas the
third term represents the two-loop contribution, where
$C_{2\mathrm{LP}}$ is a constant. We assume that $C_{2\mathrm{LP}}$
can take all possible values, and examine under what circumstances
the stable fixed point obtained in our one-loop analysis is robust.

\begin{figure}[htbp]
\center
\includegraphics[width=3.35in]{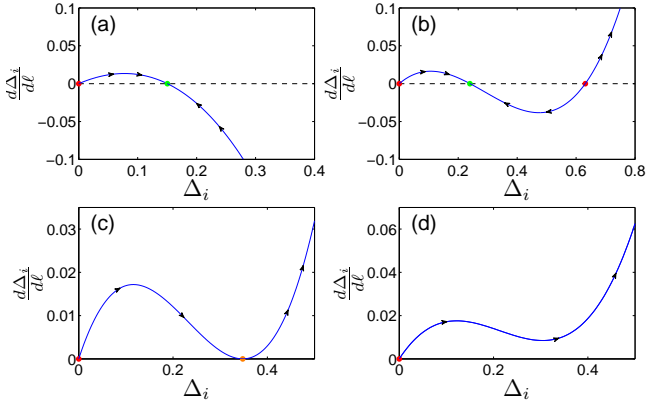}
\caption{Dependence of $\frac{d\Delta_{i}}{d\ell}$ on $\Delta_{i}$
with $i=1$ or $2$ in four different cases. (a) $C_{2\mathrm{LP}} <
0$; (b) $0 < C_{2\mathrm{LP}} < \frac{529}{192}$; (c)
$C_{2\mathrm{LP}}=\frac{529}{192}$; (d) $C_{2\mathrm{LP}} >
\frac{529}{192}$. We assume $C_{2\mathrm{LP}}=-2, 2.2, 3$ in (a),
(b), and (d), respectively. The red and green points represent
unstable and stable fixed point, respectively. The orange point in
(c) is stable from the left side, but is unstable from the right
side. \label{Fig:TWRVPXDiffTwoLoop}}
\end{figure}

There are four different cases. If $C_{2\mathrm{LP}} < 0$, there is
always a stable fixed point
\begin{eqnarray}
\Delta_{i}^{*} = \frac{\frac{23}{12} -
\sqrt{\left(\frac{23}{12}\right)^{2} -
\frac{4}{3}C_{2\mathrm{LP}}}}{2C_{2\mathrm{LP}}},
\end{eqnarray}
which is shown in Fig.~\ref{Fig:TWRVPXDiffTwoLoop}(a). If $0 <
C_{2\mathrm{LP}} < \frac{529}{192}$, as depicted in
Fig.~\ref{Fig:TWRVPXDiffTwoLoop}(b), there exists a stable fixed
point
\begin{eqnarray}
\Delta_{i}^{s*}=\frac{\frac{23}{12} -
\sqrt{\left(\frac{23}{12}\right)^{2} -
\frac{4}{3}C_{2\mathrm{LP}}}}{2C_{2\mathrm{LP}}},
\end{eqnarray}
and also a finite unstable fixed point
\begin{eqnarray}
\Delta_{i}^{us*} = \frac{\frac{23}{12} + \sqrt{\left(\frac{23}{12}
\right)^{2} - \frac{4}{3}C_{2\mathrm{LP}}}}{2C_{2\mathrm{LP}}}.
\end{eqnarray}
When $C_{2\mathrm{LP}} = \frac{529}{192}$, the above two fixed
points merge to one single fixed point
\begin{eqnarray}
\Delta_{i}^{*} = \frac{8}{23},
\end{eqnarray}
which is displayed in Fig.~\ref{Fig:TWRVPXDiffTwoLoop}(c).

\begin{figure}[htbp]
\center
\includegraphics[width=3.35in]{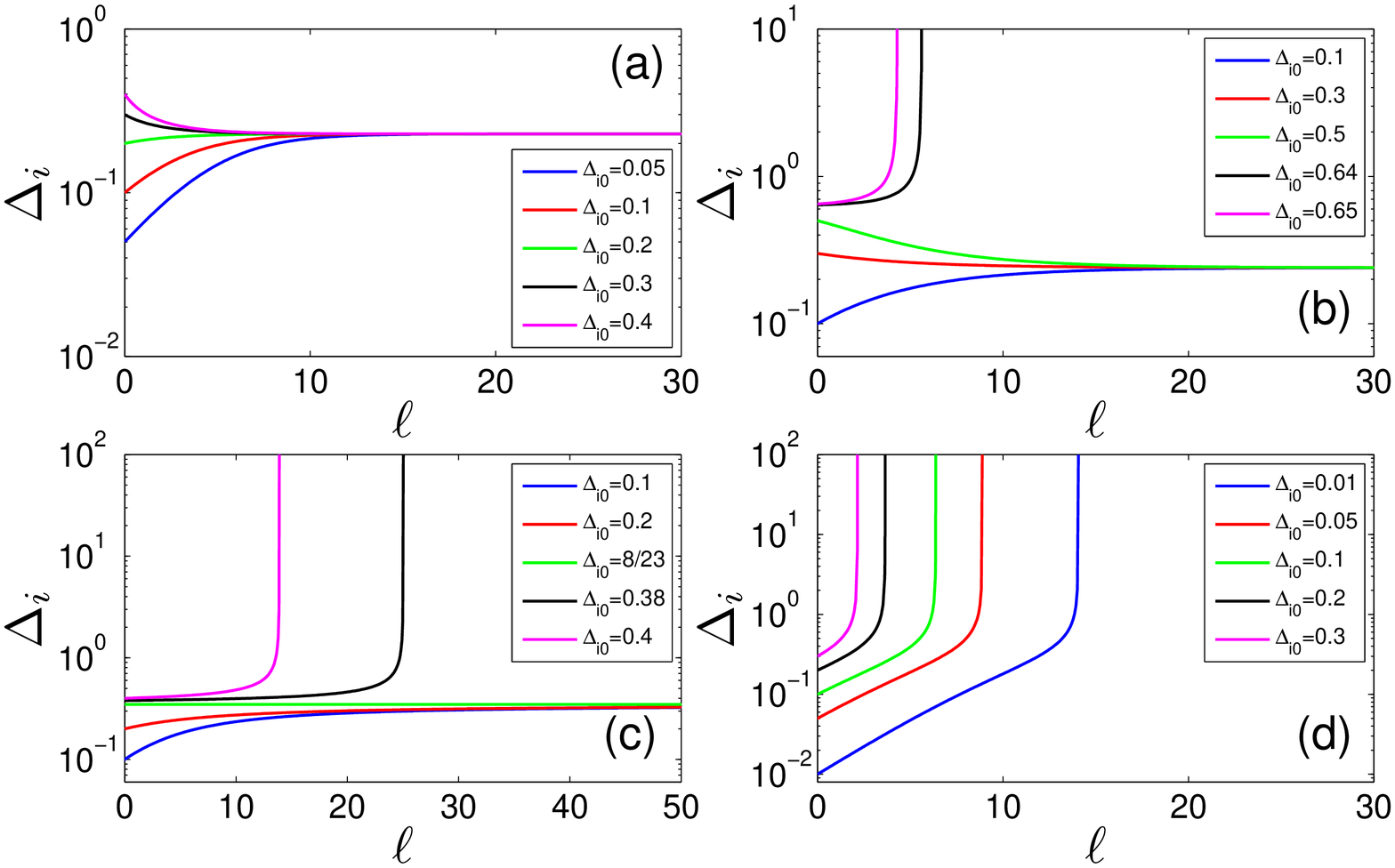}
\caption{Dependence of $\Delta_{i}$ with $i=1$ or $2$ on $\ell$ in
four different cases. We assume $C_{2\mathrm{LP}} = -2, 2.2,
\frac{529}{192}, 3$ in (a)-(d), respectively.
\label{Delta1TWTwoLoop.eps}}
\end{figure}

If the initial disorder strength $\Delta_{i0}$ is below this
critical value, i.e., $0<\Delta_{i0} < \Delta_{i}^{*}$, it always
flows to this fixed point in the lowest energy limit. However, if
$\Delta_{i0} > \Delta_{i}^{*}$, the disorder strength parameter
flows away, which turns the triple-WSM into a CDM. We finally
consider the case of $C_{2\mathrm{LP}} > \frac{529}{192}$. As shown
in Fig.~\ref{Fig:TWRVPXDiffTwoLoop}(d), there is only one unstable
fixed point $\Delta_{i}^{*} = 0$. In this case, the stable infrared
fixed point obtained in the one-loop RG analysis is eliminated by
the two-loop corrections, and even arbitrarity weak disorder drives
a CDM transition. The flows of disorder strength in triple-WSM that
contains only one component of RVP is presented in
Fig.~\ref{Delta1TWTwoLoop.eps} for four representative values of
$C_{2\mathrm{LP}}$. From Figs.~\ref{Fig:TWRVPXDiffTwoLoop} and
\ref{Delta1TWTwoLoop.eps}, we can see that there is always a stable
infrared fixed point for $C_{2\mathrm{LP}} < \frac{529}{192}$. The
concrete value of $C_{2\mathrm{LP}}$ will be calculated in the
future. We expect that numerical techniques, such as kernel
polynomial method \cite{BeraRoy16, Kobayashi14, Liu16, Pixley16A,
Sbierski16} and Lanczos method \cite{Fu17} would be employed to
determine whether the stable infrared fixed point revealed in our
one-loop RG calculation survives higher order corrections.

\subsection{Influence of Coulomb impurity on triple-WSM}

Apart from short-range disorder, there might be disorder with
long-range correlation in various SMs \cite{Fedorenko12, DasSarma11,
Nomura07, Khveshchenko07, Louvet17, Skinner14, Ominato15}. The most
frequently encountered is Coulomb impurity, which is defined in a
similar way to that of RSP, but is spatially long-ranged.
Generically, the role played by long-range disorder is more
important than short-range disorder in SMs. For 3D DSM/WSM, the
physical effects of short-range disorder and long-range disorder
turn out to be quite different. Weak short-range RSP is irrelevant
and becomes relevant if its strength is large enough. Recent RG
analysis of Louvet \emph{et al.} \cite{Louvet17} found that an
arbitrarily weak long-range disorder drives the 3D WSM to become a
CDM if such disorder decays more slowly than $1/r^{2}$ for large
$r$. It is thus clear that arbitrarily weak Coulomb impurity, which
decays as $1/r$, can lead to a SM-CDM transition. Similar conclusion
was found to be applicable to 3D DSM \cite{Skinner14, Ominato15}.

Since the present work is focused on the interplay between the
long-range Coulomb interaction and short-range disorder, we will not
present a thorough analysis of the physical effects of long-range
disorder. Here, we will only consider a special case, namely the
influence of Coulomb impurity on the low-energy properties of
triple-WSM. The extension to double-WSM and other types of
long-range disorder would be straightforward.

\begin{figure}[htbp]
\center
\includegraphics[width=3.35in]{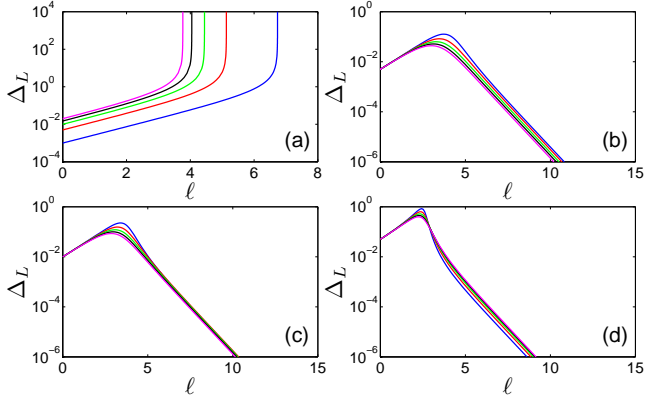}
\caption{Flow of the Coulomb impurity parameter $\Delta_{L}$ in
triple-WSM. In (a), there is only Coulomb impurity, and blue, red,
green, black, and magenta curves represent the initial values
$\Delta_{L0}=0.001, 0.005, 0.01, 0.015, 0.02$, respectively. In
(b)-(d), Coulomb interaction and Coulomb impurity are both present.
$\Delta_{L0}=0.005$, $0.01$, and $0.05$ is taken in (b), (c), and
(d), where blue, red, green, black, and magenta curves are obtained
at $\alpha_{0} = 0.005, 0.01, 0.015, 0.02, 0.025$, respectively.
Here, we set $\zeta_{0}^{t}=0.1$. \label{Fig:RGTWCoulombImpurity}}
\end{figure}

Using the replica method, one can write down the following action
for triple-Weyl fermions embedded in the potential generated by
Coulomb impurity
\begin{eqnarray}
S_{\mathrm{Coim}} &=& \frac{\Delta_{L}}{2}\int\frac{d\omega_1
d\omega_2d^{3}\mathbf{k}_{1} d^{3}\mathbf{k}_{2}
d^{3}\mathbf{k}_{3}}{(2\pi)^8} \nonumber
\\
&&\times\frac{g^{2}}{|\mathbf{k}_{1}+\mathbf{k}_{2}|_{\bot}^{2}+\eta
\left(k_{1z}+k_{2z}\right)^{2}}\nonumber
\\
&&\times\Psi^\dagger_{a}(i\omega_1,\mathbf{k}_1)\mathbbm{1}
\Psi_{a}(i\omega_1,\mathbf{k}_2)
\Psi_{b}^\dagger(i\omega_2,\mathbf{k}_3)\mathbbm{1}\nonumber
\\
&&\times\Psi_{b}(i\omega_2,-\mathbf{k}_1-\mathbf{k}_2-\mathbf{k}_3),
\end{eqnarray}
where $\Delta_{L}$ is introduced to quantify the strength of Coulomb
impurity, and $g^{2} = 4\pi e^{2}/\epsilon$ and $\eta$ are defined
in the same way as in Sec.~\ref{Sec:Model}.

To make our analysis more generic, we will consider the interplay of
Coulomb interaction, Coulomb impurity, and short-range RSP. The
coupled RG equations are
\begin{eqnarray}
\frac{dZ_{f}}{d\ell}&=&-\left(\frac{5}{4}\Delta_{0} +
C_{L1}\Delta_{L}\right)Z_{f},
\\
\frac{dB}{d\ell}&=&\left(C_{2}^{t}-\frac{5}{4}\Delta_{0} -
C_{L1}\Delta_{L}+C_{L2}\Delta_{L}\right)B,
\\
\frac{dv}{d\ell}&=&\left(C_{3}^{t}-\frac{5}{4}\Delta_{0} -
C_{L1}\Delta_{L}+C_{L3}\Delta_{L} \right)v,
\\
\frac{d\alpha}{d\ell} &=& \Big(-C_{\bot}^{t}-C_{3}^{t} +
\frac{5}{4}\Delta_{0} \nonumber
\\
&& +C_{L1}\Delta_{L} - C_{L3}\Delta_{L}\Big)\alpha,
\\
\frac{d\beta^{t}}{d\ell} &=& \left(\frac{4}{3}-\frac{2}{3}C_{2}^{t}
+ C_{3}^{t}-\frac{5}{12}\Delta_{0} -
\frac{1}{3}C_{L1}\Delta_{L}\right.
\nonumber \\
&&\left.-\frac{2}{3}C_{L2}\Delta_{L}+C_{L3}\Delta_{L} -
\beta^{t}\right)\beta^{t},
\\
\frac{d\eta}{d\ell}&=&\left(-\frac{4}{3}-C_{\bot}^{t} +
\beta^{t}\right)\eta,
\\
\frac{d\Delta_{0}}{d\ell} &=&
\frac{1}{3}\Delta_{0}-\left(\frac{2}{3}C_{2}^{t}+C_{3}^{t}+2
C_{\bot}^{t} +2\beta^{t}\right)\Delta_{0}\nonumber
\\
&&+\left(\frac{5}{3}C_{L1}-\frac{2}{3}C_{L2}-C_{L3}
\right)\Delta_{0}\Delta_{L}\nonumber
\\
&&+\frac{25}{12}\Delta_{0}^{2}, \label{Eq:RGTWCIDelta0}
\\
\frac{d\Delta_{L}}{d\ell}
&=&\Delta_{L}-\left(C_{3}^{t}+3C_{\bot}^{t}
+2\beta^{t}\right)\Delta_{L}+\frac{5}{4}\Delta_{0}\Delta_{L}\nonumber
\\
&&+\left(C_{L1}-C_{L3}\right)\Delta_{L}^{2}. \label{Eq:RGTWCIDeltaL}
\end{eqnarray}
The parameter $\Delta_{L}$ has been redefined as follows:
\begin{eqnarray}
\frac{c_{f}\Delta_{L}g^{2}}{v\Lambda} \rightarrow \Delta_{L}.
\end{eqnarray}
The concrete expressions for $C_{Li}\equiv C_{Li}(\zeta^{t})$, where
$i=1, 2, 3$ and $\zeta^{t} = \frac{\eta B^{\frac{2}{3}}
\Lambda^{\frac{4}{3}}}{v^{2}}$, are presented in
Appendix~\ref{App:CLiLRDisorder}.

If there is only Coulomb impurity, the dependence of $\Delta_{L}$ on
$\ell$ is displayed in Fig.~\ref{Fig:RGTWCoulombImpurity}(a). The
parameter $\Delta_{L}$ always flows to the strong coupling regime,
thus the Coulomb impurity always leads to a CDM phase. When both
Coulomb impurity and long-range Coulomb interaction are considered,
the parameter $\Delta_{L}$ exhibits distinct behavior. As can be
seen from Figs.~\ref{Fig:RGTWCoulombImpurity}(b)-(d), $\Delta_{L}$
flows to zero rapidly at low energies. Thus, the SM phase is
restored due to long-range Coulomb interaction, and the Coulomb
impurity becomes relatively unimportant. This behavior is presumably
caused by the special anisotropic screening effect induced by
Coulomb interaction.

\begin{figure}[htbp]
\center
\includegraphics[width=3.35in]{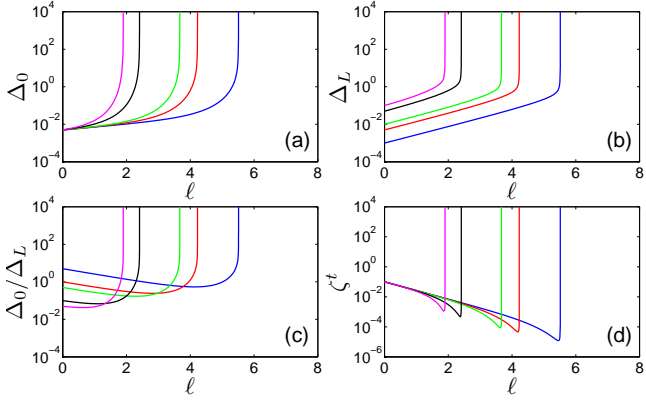}
\caption{Flow of $\Delta_{0}$ in (a), $\Delta_{L}$ in (b),
$\Delta_{0}/\Delta_{L}$ in (c), and $\zeta^{t}$ in (d). Both RSP and
Coulomb impurity are considered, but Coulomb interaction is ignored.
$\Delta_{00}=0.005$ is taken, and blue, red, green, black, and
magenta curves represent the initial values $\Delta_{L0}=0.001,
0.005, 0.01, 0.05, 0.1$, respectively. Here, we choose an initial
value $\zeta_{0}^{t}=0.1$. \label{Fig:RGTWCoulombImpurityRSP}}
\end{figure}

We then neglect the Coulomb interaction and analyze the interplay
between Coulomb impurity and short-range RSP. The RG equations for
$\Delta_{0}$ and $\Delta_{L}$ are
\begin{eqnarray}
\frac{d\Delta_{0}}{d\ell} &=&\left[
\frac{1}{3}+\left(\frac{5}{3}C_{L1}-\frac{2}{3}C_{L2}-C_{L3}
\right)\Delta_{L}\right.\nonumber
\\
&&\left.+\frac{25}{12}\Delta_{0}\right]\Delta_{0},
\label{Eq:TWCoulombImpuRSPA}
\\
\frac{d\Delta_{L}}{d\ell} &=& \left[1+\frac{5}{4}\Delta_{0} +
\left(C_{L1}-C_{L3}\right)\Delta_{L}\right]\Delta_{L}.
\label{Eq:TWCoulombImpuRSPB}
\end{eqnarray}
In this case, the RG flows of $\Delta_{0}$, $\Delta_{L}$, and
$\Delta_{0}/\Delta_{L}$ are presented in
Figs.~\ref{Fig:RGTWCoulombImpurityRSP}(a)-(c) respectively. We
observe that $\Delta_{0}$, $\Delta_{L}$, and $\Delta_{0}/\Delta_{L}$
all formally diverge at some finite energy scale. Thus, RSP
dominates over Coulomb impurity. The parameter $\zeta^{t}$ also
flows to infinity simultaneously at finite energy scale, as shown in
Fig.~\ref{Fig:RGTWCoulombImpurity}(d).

\begin{figure}[htbp]
\center
\includegraphics[width=3.35in]{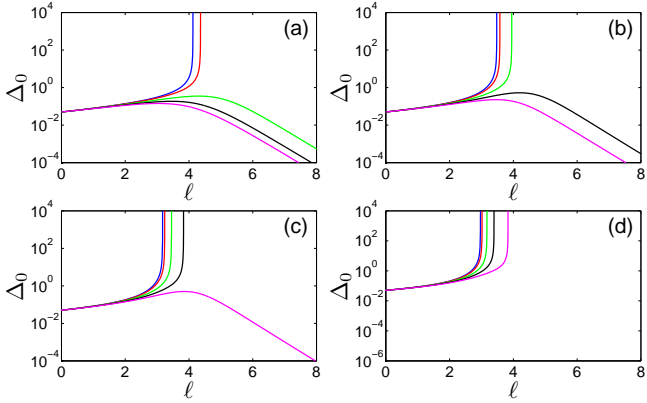}
\caption{Flow of $\Delta_{0}$ caused by the interplay of RSP,
Coulomb impurity, and Coulomb interaction in triple-WSM. Blue, red,
green, black, and magenta curves represent the initial values
$\alpha_{0}=0.003, 0.005, 0.01, 0.015, 0.02$, respectively.
$\Delta_{L0}$ is taken as $0.001$, $0.004$, $0.007$, and $0.01$ in
(a), (b), (c), and (d) respectively. An initial value
$\zeta_{0}^{t}=0.1$ is chosen. \label{Fig:RGTWCoulombImpurityAll}}
\end{figure}

To understand these results, we now analyze the running behavior of
the ratio $\Delta_{L}/\Delta_{0}$. The RG equation is
\begin{eqnarray}
\frac{d(\Delta_{L}/\Delta_{0})}{d\ell} = \left(\frac{2}{3} -
\frac{10}{12}\Delta_{0}-\frac{2}{3}C_{L1}+\frac{2}{3}C_{L2}\right)
\frac{\Delta_{L}}{\Delta_{0}}.
\end{eqnarray}
The first term in the parenthesis, namely $\frac{2}{3}$, tells us
that Coulomb impurity dominates over RSP at the tree-level. The
second term is negative, and formally diverges as $\Delta_{0}$ flows
to infinity. The contribution $\frac{2}{3}(C_{L2}-C_{L1})$ decreases
with growing $\zeta^{t}$. It is easy to verify that the summation of
all the terms in the parenthesis goes to negative infinity, which
implies that the ratio $\Delta_{L}/\Delta_{0}$ eventually flows to
zero. This explains why RSP becomes more important than the Coulomb
impurity at low energies. The apparently anisotropic dispersion of
triple-Weyl fermions is likely responsible for this property.

\begin{figure}[htbp]
\center
\includegraphics[width=3.35in]{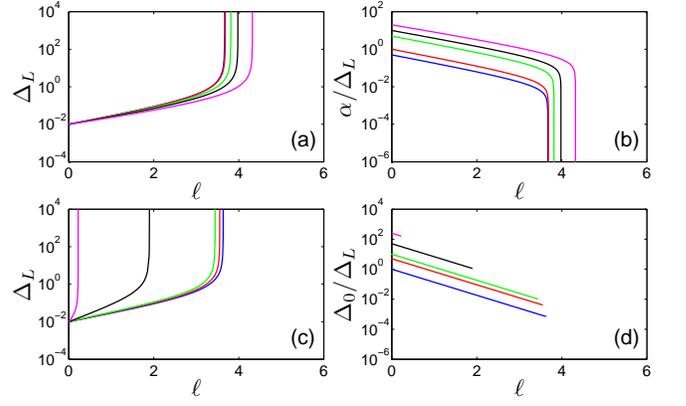}
\caption{RG results for usual WSM. Coulomb impurity and Coulomb
interaction are both included in (a) and (b) with
$\Delta_{L0}=0.01$, where blue, red, green, black, and magenta
curves stand for $\alpha_{0}=0.005, 0.01, 0.05, 0.1, 0.2$,
respectively. Coulomb impurity and RSP are both considered in (c)
and (d) with $\Delta_{L0}=0.01$, where blue, red, green, black, and
magenta curves represent $\Delta_{0}=0.01, 0.05, 0.1, 0.5, 2.2$,
respectively.\label{Fig:RGWSMCoulombImpu}}
\end{figure}

For triple-WSM that contains RSP, Coulomb impurity, and Coulomb
interaction, the $\ell$-dependence of $\Delta_{0}$ is shown in
Fig.~\ref{Fig:RGTWCoulombImpurityAll}. We observe that, for
different initial conditions, $\Delta_{0}$ either vanishes or flows
to infinity, which indicates that triple-WSM could be in the SM or
CDM phase. Comparing the results presented in
Fig.~\ref{Fig:RGTWCoulombImpurityAll}, one finds that increasing the
strength of Coulomb impurity promotes CDM transition. If the
strength parameters for RSP and Coulomb impurity are fixed, the SM
phase can be restored by strong Coulomb interaction.

We now consider the impact of Coulomb impurity in the usual WSM,
which will be compared with the case of triple-WSM. In the presence
of Coulomb interaction, Coulomb impurity, and RSP, the RG equations
for the corresponding parameters are given by
\begin{eqnarray}
\frac{dZ_{f}}{d\ell}&=&-\left(\Delta_{0}+\Delta_{L}\right)Z_{f},
\\
\frac{dv}{d\ell}&=&\left(\frac{2\alpha}{3\pi}-\Delta_{0} -
\frac{4}{3}\Delta_{L}\right)v,
\\
\frac{d\alpha}{d\ell} &=& \alpha \left(\Delta_{0} +
\frac{4}{3}\Delta_{L} - \frac{4}{3\pi}\alpha\right),
\\
\frac{d\Delta_{0}}{d\ell} &=& \left(-1+2\Delta_{0}+\frac{8}{3}
\Delta_{L}-\frac{8}{3\pi}\alpha\right)\Delta_{0},
\\
\frac{d\Delta_{L}}{d\ell} &=& \left(1+\frac{8}{3}\Delta_{L} +
2\Delta_{0}-\frac{10}{3\pi}\alpha\right)\Delta_{L},
\end{eqnarray}
where we have made the replacements:
\begin{eqnarray}
\frac{\Delta_{0}\Lambda}{2\pi^2v^2}\rightarrow \Delta_{0}, \quad
\frac{\Delta_{L}g^{2}}{2\pi^{2}v^{2}\Lambda} \rightarrow \Delta_{L}.
\end{eqnarray}

We present in in Fig.~\ref{Fig:RGWSMCoulombImpu}(a) the results
obtained when there are Coulomb impurity and Coulomb interaction. We
find that $\Delta_{L}$ exhibits runaway behavior, and that $\alpha$
also flows to strong coupling. However, the ratio
$\alpha/\Delta_{L}$ vanishes rapidly with growing $\ell$, as clearly
shown in Fig.~\ref{Fig:RGWSMCoulombImpu}(b). Therefore, the usual
WSM undergoes a CDM phase transition driven by Coulomb impurity.

We then assume that Coulomb impurity and RSP exist simultaneously,
but the long-range Coulomb interaction is neglected. The RG
equations for $\Delta_{0}$ and $\Delta_{L}$ are
\begin{eqnarray}
\frac{d\Delta_{0}}{d\ell} &=& \left(-1+2\Delta_{0} +
\frac{8}{3}\Delta_{L}\right)\Delta_{0},\label{Eq:UTWCoulombImpRSPA}
\\
\frac{d\Delta_{L}}{d\ell} &=& \left(1+\frac{8}{3}\Delta_{L} +
2\Delta_{0}\right)\Delta_{L}. \label{Eq:UTWCoulombImpRSPB}
\end{eqnarray}
As shown in Fig.~\ref{Fig:RGWSMCoulombImpu}(c), $\Delta_{L}$ always
approaches to infinity at some finite energy scale. $\Delta_{0}$
also flows to strong coupling at the same energy scale. As depicted
in Fig.~\ref{Fig:RGWSMCoulombImpu}(d), $\Delta_{0}/\Delta_{L}$
deceases monotonously with growing of $\ell$. If the initial value
$\Delta_{00}/\Delta_{L0}$ is very large, $\Delta_{0}/\Delta_{L}$
still takes a large value at the finite energy scale, in which
$\Delta_{0}$ and $\Delta_{L}$ become divergent. If the initial value
$\Delta_{00}/\Delta_{L0}$ is not very large, $\Delta_{0}/\Delta_{L}$
takes a value smaller than unity finally. Form the
Eqs.~(\ref{Eq:UTWCoulombImpRSPA}) and (\ref{Eq:UTWCoulombImpRSPB}),
we can get the RG equation for the ratio $\Delta_{L}/\Delta_{0}$
\begin{eqnarray}
\frac{d(\Delta_{L}/\Delta_{0})}{d\ell}=2\frac{\Delta_{L}}{\Delta_{0}}.
\end{eqnarray}
It is clear that there is only tree-level contribution. The one-loop
order corrections do not alter the behavior of
$\Delta_{L}/\Delta_{0}$. If Coulomb impurity, RSP, and Coulomb
interaction are all included, usual WSM is always turned into the
CDM phase.

The above analysis show that the triple-WSM and usual WSM exhibit
quite different behavior in response to the interplay of Coulomb
impurity, RSP, and Coulomb interaction, which should be attributed
to the difference in their fermion dispersions.

\section{Summary\label{Sec:Summary}}

In summary, we have systematically studied the low-energy behavior
of double- and triple-WSMs induced by the interplay between
long-range Coulomb interaction and disorder. After performing a
detailed RG analysis, we have showed that such an interplay has
distinct influences on the dynamics of double- and triple-Weyl
fermions. The double-WSM is always in a CDM phase if the system
contains any type of disorder, such as RSP or RVP, and this feature
is not altered by the addition of Coulomb interaction. However, the
low-energy behavior of triple-Weyl fermions depend crucially on the
type and strength of disorder. In the non-interacting limit, either
RSP or $z$-RVP leads to a CDM transition, and $x$-RVP or $y$-RVP
results in a stable quantum critical state. The interplay of RSP and
weak Coulomb interaction turns the triple-WSM into a CDM phase, but
the interplay of RSP and strong Coulomb interaction renders the
stability of SM state. When the triple-WSM contains both $x$-RVP, or
$y$-RVP, and Coulomb interaction, the system always flows to a
stable infrared fixed point. The stability of this fixed point
against two-loop corrections is also discussed. However, this
problem is only partly answered, and more elaborate RG calculations
are required to completely solve the problem. Finally, the interplay
of $z$-RVP and Coulomb interaction may drive a QPT between SM and
Mott insulator. We have demonstrated in great detail that the marked
difference between the low-energy properties of double- and
triple-WSMs is owing to the distinct response of the Hamiltonian
under the transformation $\mathbf{k}\rightarrow -\mathbf{k}$.

We have also considered the impact of long-range Coulomb impurity on
the low-energy behavior of triple-WSM. After making a RG analysis of
the complicated interplay of Coulomb impurity, Coulomb interaction,
and RSP, we find that, while the Coulomb impurity always drives the
system to become CDM, the Coulomb interaction can effectively
suppress the role played by Coulomb impurity and protect the SM
state.

The diverse phases and the transitions between them predicted by our
RG analysis could be verified by performing angle-resolved
photoemission spectroscopy (ARPES) \cite{Damascelli03, Hasan17} and
transport measurements. Recent first-principle calculations
suggested that HgCr$_{2}$Se$_{4}$ \cite{Fang12} and SrSi$_{2}$
\cite{Huang16} are two promising candidates of the double-WSM. In
addition, special SM systems, in which the fermions exhibit a
linear dependence on one momentum component and cubic dependence on
other two components, were predicted to be realizable in
Rb(MoTe)$_{3}$ and Tl(MoTe)$_{3}$ \cite{LiuZunger17}. It is also
possible to prepare the multi-WSM materials by microwave experiments
\cite{ChenChan16, ChangChan17}. We expect that our theoretical
predictions would be verified in the aforementioned materials in the
future.

\section*{ACKNOWLEDGEMENTS}

We would like to acknowledge the support by the Ministry of Science
and Technology of China under Grants 2016YFA0300404 and
2017YFA0403600, and the support by the National Natural Science
Foundation of China under Grants 11574285, 11504379, 11674327, and
U1532267. J.R.W. is also supported by the Natural Science Foundation
of Anhui Province under Grant 1608085MA19.

\appendix

\section{Propagators}

We now present the propagators of double- and triple-Weyl fermions
and bosonic field that is introduced to represent the long-range
Coulomb interaction. The boson self-energy is calculated in
Appendix~\ref{App:BosonSelfEnergy}. We then give the fermion
self-energy induced by Coulomb interaction and disorder scattering
in Appendix~\ref{App:FermiSelfEnergy}. In
Appendix~\ref{App:CoulombVertexCorrection}, the corrections to the
fermion-boson coupling are computed. The vertex corrections to the
fermion-disorder couplings are calculated in
Appendix~\ref{App:DisVertexCorrection}. The RG equations for the
model parameters of double- and triple-WSMs are derived in
Appendix~\ref{App:RGEquations}. The expressions of $C_{Li}$, which
enter into the RG equations for Coulomb impurity, are shown in
Appendix~\ref{App:CLiLRDisorder}.

The free propagator of double-Weyl fermions is \cite{Lai15, Jian15}
\begin{eqnarray}
G_{0}^{d}(\omega,\mathbf{k}) = \frac{1}{-i\omega+
Ad_{1}(\mathbf{k})\sigma_{1} + Ad_{2}(\mathbf{k})\sigma_{2} +
vk_{z}\sigma_{3}}, \label{Eq:FermionPropagatorDW}
\end{eqnarray}
where $d_{1}(\mathbf{k})=k_{x}^{2}-k_{y}^{2}$ and
$d_{2}(\mathbf{k})=2k_{x}k_{y}$. The free propagator of triple-Weyl
fermions can be written as  \cite{Zhang16}
\begin{eqnarray}
G_{0}^{t}(\omega,\mathbf{k}) =
\frac{1}{-i\omega+Bg_{1}(\mathbf{k}) \sigma_{1} +
Bg_{2}(\mathbf{k})\sigma_{2} + vk_{z}\sigma_{3}},
\label{Eq:FermionPropagatorTW}
\end{eqnarray}
where $g_{1}(\mathbf{k})=k_{x}^{3}-3k_{x}k_{y}^{2}$ and
$g_{2}(\mathbf{k})=k_{y}^{3}-3k_{y}k_{x}^{2}$. The propagator of
bosonic field $\phi$ reads
\begin{eqnarray}
D_{0}(\Omega,\mathbf{q}) = \frac{1}{q_{\bot}^{2}+\eta q_{z}^{2}}.
\label{Eq:BosonPropagator}
\end{eqnarray}

\begin{figure}[htbp]
\center
\includegraphics[width=1.8in]{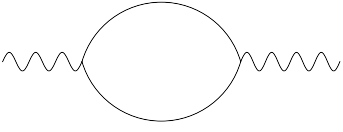}
\caption{Self-energy of bosonic field. The solid line represents the
free fermion propagator, and the wavy line represents the boson
propagator that is equivalent to the Coulomb interaction function.
\label{Fig:BosonSelfEnergy}}
\end{figure}

\section{Boson self-energy \label{App:BosonSelfEnergy}}

As shown in Fig.~\ref{Fig:BosonSelfEnergy}, to the leading order of
perturbative expansion, the self-energy of bosonic field $\phi$ is
given by
\begin{eqnarray}
\Pi^{d,t}(\Omega,\mathbf{q}) &=& -g^{2}\int \frac{d\omega}{2\pi} \int
\frac{d^3\mathbf{k}}{(2\pi)^{3}}
\mathrm{Tr}\left[G_{0}^{d,t}(\omega,\mathbf{k})\right.\nonumber
\\
&&\left.\times G_{0}^{d,t}
\left(\omega+\Omega,\mathbf{k} +
\mathbf{q}\right)\right].\label{Eq:SelfEnergyBoson}
\end{eqnarray}

\subsection{Double-Weyl fermions}

Substituting Eq.~(\ref{Eq:FermionPropagatorDW}) into
Eq.~(\ref{Eq:SelfEnergyBoson}), and then taking the limit $\Omega =
0$, we obtain
\begin{eqnarray}
\Pi^{d}(0,\mathbf{q}) &=& 2g^{2}\int_{-\infty}^{+\infty}
\frac{d\omega}{2\pi}\int'\frac{d^3\mathbf{k}}{(2\pi)^{3}}\nonumber
\\
&&\times\frac{1}{\left(\omega^{2} + E_{\mathbf{k}}^{2}\right)
\left(\omega^{2} + E_{\mathbf{k}+\mathbf{q}}^{2}\right)}
\Big\{\omega^{2}\nonumber \\
&&-A^{2}\left(k_{x}^{2}-k_{y}^{2}\right)
\left[\left(k_{x}+q_{x}\right)^{2} -
\left(k_{y}+q_{y}\right)^{2}\right]\nonumber
\\
&&-4A^{2}k_{x}k_{y}\left(k_{x}+q_{x}\right)
\left(k_{y}+q_{y}\right)\nonumber
\\
&&-v^{2}k_{z}\left(k_{z}+q_{z}\right)\Big\},
\end{eqnarray}
where $E_{\mathbf{k}}=\sqrt{A^2k_{\bot}^{4}+v^2k_{z}^{2}}$.
Expanding $q_{i}$ up to the quadratical order leads us to
\begin{eqnarray}
\Pi^{d}(0,\mathbf{q}) &\approx& q_{\bot}^{2}\frac{g^{2}}{4\pi^{2}}
\int'dk_{\bot}d|k_{z}|k_{\bot}\left(\frac{2A^{2}
k_{\bot}^{2}}{E_{\mathbf{k}}^{3}} -
\frac{A^{4}k_{\bot}^{6}}{E_{\mathbf{k}}^{5}}\right)\nonumber
\\
&&+q_{z}^{2}\frac{g^{2}}{8\pi^{2}}\int'dk_{\bot}d|k_{z}|k_{\bot}
\frac{v^{2}A^{2}k_{\bot}^{4}}{E_{\mathbf{k}}^{5}}.
\end{eqnarray}
To proceed, we introduce two new variables
\begin{eqnarray}
E = \sqrt{A^2k_{\bot}^{4}+v^{2}k_{z}^{2}},\qquad \delta =
\frac{Ak_{\bot}^{2}}{v\left|k_{z}\right|},
\label{Eq:TrasformationIntegerationDWA}
\end{eqnarray}
and perform integrations over $E$ and $\delta$ within the ranges
$b\Lambda < E < \Lambda$ with $b = e^{-\ell}$ and $0 < \delta <
\infty$, which leads to
\begin{eqnarray}
\Pi^{d}(0,\mathbf{q}) = q_{\bot}^{2}C_{\bot}^{d}\ell +
q_{z}^{2}C_{z}^{d}\ell,
\end{eqnarray}
where
\begin{eqnarray}
C_{\bot}^{d} = \frac{g^{2}}{6\pi^{2}v},\qquad C_{z}^{d} =
\frac{g^{2}}{64\pi v} \frac{v^{2}}{A\Lambda}.
\end{eqnarray}

\subsection{Triple-Weyl fermions}

Substituting Eq.~(\ref{Eq:FermionPropagatorTW}) into
(\ref{Eq:SelfEnergyBoson}) and taking the limit $\Omega=0$, we find
that the boson self-energy has the form
\begin{eqnarray}
\Pi^{t}(0,\mathbf{q}) &=& 2g^{2}\int_{-\infty}^{+\infty}
\frac{d\omega}{2\pi}\int'\frac{d^3\mathbf{k}}{(2\pi)^{3}}\nonumber
\\
&&\times\frac{1}{\left(\omega^{2}+E_{\mathbf{k}}^{2}\right)
\left(\omega^{2}+E_{\mathbf{k}+\mathbf{q}}^{2}\right)}
\Big\{\omega^{2}\nonumber
\\
&&-B^{2}\left(k_{x}^{3}-3k_{x}k_{y}^{2}\right)
\left[\left(k_{x}+q_{x}\right)^{3} -
3\left(k_{x}+q_{x}\right)\right.\nonumber
\\
&&\left.\times\left(k_{y}+q_{y}\right)^{2} \right] -
B^{2}\left(k_{y}^{3}-3k_{y}k_{x}^{2}\right)\nonumber
\\
&&\times\left[\left(k_{y} +
q_{y}\right)^{3} - 3\left(k_{y}+q_{y}\right)\left(k_{x} +
q_{x}\right)^{2} \right] \nonumber
\\
&&- v^2k_{z}\left(k_{z}+q_{z}\right)\Big\},
\label{Eq:SelfEnergyBosonTWB}
\end{eqnarray}
where $E_{\mathbf{k}}=\sqrt{B^2k_{\bot}^{6}+v^2k_{z}^{2}}$.
Expanding to the quadratical order of $q_{i}$, we get
\begin{eqnarray}
\Pi^{t}(0,\mathbf{q}) &=& q_{\bot}^{2} \frac{9g^{2}}{16\pi^{2}}
\int'dk_{\bot}d|k_{z}|k_{\bot}
\left(\frac{2B^{2}k_{\bot}^{4}}{E_{\mathbf{k}}^{3}} -
\frac{B^{4}k_{\bot}^{10}}{E_{\mathbf{k}}^{5}}\right)\nonumber
\\
&&+q_{z}^{2}\frac{g^{2}}{16\pi^{2}}\int'dk_{\bot}d|k_{z}|k_{\bot}
\frac{v^2 B^2k_{\bot}^{6}}{E_{\mathbf{k}}^{5}}.
\label{Eq:SelfEnergyBosonTWC}
\end{eqnarray}
Utilizing the transformations
\begin{eqnarray}
E = \sqrt{B^2k_{\bot}^{6}+v^{2}k_{z}^{2}},\qquad \delta =
\frac{Bk_{\bot}^{3}}{v\left|k_{z}\right|},
\label{Eq:TrasformationIntegerationTWA}
\end{eqnarray}
and carrying out the integrations of $E$ and $\delta$, we obtain
\begin{eqnarray}
\Pi^{t}(0,\mathbf{q}) = q_{\bot}^{2} C_{\bot}^{t}\ell +
q_{z}^{2}C_{z}^{t}\ell,
\end{eqnarray}
where
\begin{eqnarray}
C_{\bot}^{t}=\frac{g^{2}}{4\pi^{2}v},\qquad C_{z}^{t} =
\frac{\Gamma\left(\frac{1}{3}\right)g^{2}}{120\pi^{\frac{3}{2}}
\Gamma\left(\frac{5}{6}\right)v}\frac{v^2}{B^{\frac{2}{3}}
\Lambda^{\frac{4}{3}}}.
\end{eqnarray}

\section{Fermion self-energy corrections \label{App:FermiSelfEnergy}}

We now compute the fermion self-energy corrections caused by Coulomb
interaction and disorder.

\subsection{Fermion self-energy due to Coulomb interaction}

As displayed in Fig.~\ref{Fig:FermionSelfEnergy}(a), the fermion
self-energy caused by Coulomb interaction is
\begin{eqnarray}
\Sigma_{C}^{d,t}(\omega,\mathbf{k}) &=&-g^{2}
\int'\frac{d\Omega}{2\pi}\frac{d^{3}\mathbf{q}}{(2\pi)^{3}}
G_{0}^{d,t}(\omega+\Omega,\mathbf{k}+\mathbf{q})\nonumber
\\
&&\times D_{0}(\Omega,\mathbf{q}).
\label{Eq:SelfEnergyFermionCoulomb}
\end{eqnarray}

\subsubsection{Double-Weyl fermions\label{App:SelfEnergyCoulombDW}}

Substituting Eqs.~(\ref{Eq:FermionPropagatorDW}) and
(\ref{Eq:BosonPropagator}) into
Eq.~(\ref{Eq:SelfEnergyFermionCoulomb}),
$\Sigma_{C}^{d}(\omega,\mathbf{k})$ can be approximated as
\begin{eqnarray}
\Sigma_{C}^{d}(\omega,\mathbf{k}) &\approx& \left\{i\omega C_{1}^{d}
- A\left[d_{1}(\mathbf{k})\sigma_{1} +
d_{2}(\mathbf{k})\sigma_{2}\right]C_{2}^{d} \right.\nonumber
\\
&&\left.-vk_{z}\sigma_{3}C_{3}^{d}\right\}\ell.
\label{Eq:DWSelfEnergyResultMainText}
\end{eqnarray}
where $C_{i}^{d}\equiv C_{i}^{d}(\alpha,\zeta^{d})$ with
\begin{eqnarray}
\alpha=\frac{e^{2}}{v\epsilon}, \quad \zeta^{d}=\frac{\eta
A\Lambda}{v^{2}}. \label{Eq:DefZetad}
\end{eqnarray}
The expressions of $C_{1}^{d}$, $C_{2}^{d}$ and $C_{3}^{d}$ take
the form
\begin{eqnarray}
C_{1}^{d} &=& 0, \label{Eq:C1ExpressionDW}
\\
C_{2}^{d} &=& \frac{\alpha}{4\pi}\int_{0}^{+\infty}d\delta
\frac{1}{\left(1+\delta^2\right)}\left[2-3\delta^{2}
+3\frac{\delta^{4}}{\left(1+\delta^2\right)}\right]\nonumber
\\
&&\times\frac{1}{\delta\left(1+\delta^2\right)^{\frac{1}{2}} +
\zeta^{d}}, \label{Eq:C2ExpressionDW}
\\
C_{3}^{d}&=&\frac{\alpha}{2\pi}\int_{0}^{+\infty}d\delta
\frac{\delta^{2}}{\left(1+\delta^2\right)}
\frac{1}{\delta\left(1+\delta^2\right)^{\frac{1}{2}} + \zeta^{d}}.
\label{Eq:C3ExpressionDW}
\end{eqnarray}

\subsubsection{Tripe-Weyl fermions\label{Sec:SelfEnergyCoulombTW}}

Substituting Eqs.~(\ref{Eq:FermionPropagatorTW}) and
(\ref{Eq:BosonPropagator}) into
Eq.~(\ref{Eq:SelfEnergyFermionCoulomb}),
$\Sigma_{C}^{t}(\omega,\mathbf{k})$ can be approximately written as
\begin{eqnarray}
\Sigma_{C}^{t}(\omega,\mathbf{k}) &\approx& \left\{i\omega C_{1}^{t}
- B\left[g_{1}(\mathbf{k})\sigma_{1} +
g_{2}(\mathbf{k})\sigma_{2}\right]C_{2}^{t}\right.\nonumber
\\
&&\left.-vk_{z}\sigma_{3}C_{3}^{t}\right\}\ell,
\label{Eq:TWSelfEnergyResultMainText}
\end{eqnarray}
where $C_{i}^{t}\equiv C_{i}^{t}(\alpha,\zeta^{t})$ with
\begin{eqnarray}
\zeta^{t}=\frac{\eta B^{\frac{2}{3}}\Lambda^{\frac{4}{3}}}{v^2}.
\label{Eq:DefBetatGammat}
\end{eqnarray}
The expressions of $C_{1}^{t}$, $C_{2}^{t}$, and $C_{3}^{t}$ are
given by
\begin{eqnarray}
C_{1}^{t}&=&0, \label{Eq:C1ExpressionTW}
\\
C_{2}^{t}&=&\frac{\alpha}{6\pi} \int_{0}^{+\infty}d\delta
\frac{1}{\delta^{\frac{1}{3}}\left(1+\delta^{2}\right)^{\frac{5}{6}}}
\left[2-17\delta^{2}+\frac{135}{4}
\frac{\delta^{4}}{1+\delta^{2}}\right.\nonumber
\\
&&\left. -\frac{135}{8}\frac{\delta^{6}}{\left(1 +
\delta^{2}\right)^{2}}\right]\frac{1}{\delta^{\frac{2}{3}}
\left(1+\delta^{2}\right)^{\frac{2}{3}}+\zeta^{t}},
\label{Eq:C2ExpressionTW}
\\
C_{3}^{t}&=&\frac{\alpha}{3\pi} \int_{0}^{+\infty} d\delta
\frac{\delta^{\frac{5}{3}}}{\left(1+\delta^{2}\right)^{\frac{5}{6}}}
\frac{1}{\delta^{\frac{2}{3}}
\left(1+\delta^{2}\right)^{\frac{2}{3}}+\zeta^{t}}.
\label{Eq:C3ExpressionTW}
\end{eqnarray}

\begin{figure}[htbp]
\center
\includegraphics[width=3.3in]{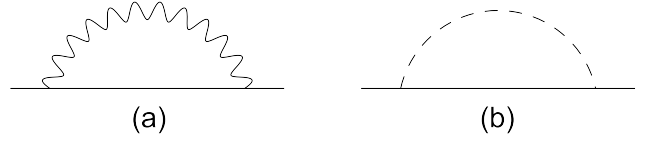}
\caption{Self-energy of fermions due to (a) Coulomb interaction and
(b) disorder. The dashed line represents disorder scattering.
\label{Fig:FermionSelfEnergy}}
\end{figure}

\subsection{Fermion self-energy due to disorder scattering
\label{App:SelfEnergyDis}}

According to Fig.~\ref{Fig:FermionSelfEnergy}(b), the self-energy of
fermions leaded by the disorder scattering takes the form
\begin{eqnarray}
\Sigma_{\mathrm{dis}}^{d,t}(\omega) = \sum_{j=0}^{3}\Delta_{j}\int'
\frac{d^3\mathbf{k}}{(2\pi)^{3}}\Gamma_{j}
G_{0}^{d,t}(\omega,\mathbf{k})\Gamma_{j}.
\label{Eq:SelfEnergyFermionDis}
\end{eqnarray}

\subsubsection{Double-Weyl fermions \label{App:SelfEnergyDisDW}}

Substituting Eq.~(\ref{Eq:FermionPropagatorDW}) into
(\ref{Eq:SelfEnergyFermionDis}), and employing the transformations
(\ref{Eq:TrasformationIntegerationDWA}), $\Sigma_{dis}^{d}$ can be
calculated as follows
\begin{eqnarray}
\Sigma_{\mathrm{dis}}^{d}(\omega) &=& i\omega\sum_{j=0}^{3}
\frac{\Delta_{j}}{4\pi^{2}vA}\int_{b\Lambda}^{\Lambda}dE
\frac{1}{E}\int_{0}^{+\infty}d\delta \frac{1}{1+\delta^2}\nonumber
\\
&=&i\omega\sum_{j=0}^{3}\frac{\Delta_{j}}{8\pi vA}\ell.
\end{eqnarray}

\subsubsection{Triple-Weyl fermions \label{App:SelfEnergyDisTW}}

Substituting Eq.~(\ref{Eq:FermionPropagatorTW}) into
Eq.~(\ref{Eq:SelfEnergyFermionDis}), we obtain
\begin{eqnarray}
\Sigma_{\mathrm{dis}}^{t}(\omega) &=&
i\omega\sum_{j=0}^{3}\frac{\Delta_{j}}{6\pi^{2}vB^{\frac{2}{3}}}
\int_{b\Lambda}^{\Lambda}\frac{1}{E^{\frac{4}{3}}} dE
\int_{0}^{+\infty}d\delta\nonumber
\\
&&\times\frac{1}{\delta^{\frac{1}{3}}
\left(1+\delta^{2}\right)^\frac{5}{6}} \nonumber
\\
&&\approx
i\omega\sum_{j=0}^{3}\frac{5}{4}\frac{\Delta_{j}c_{f}}{vB^{\frac{2}{3}}
\Lambda^{\frac{1}{3}}}\ell,
\end{eqnarray}
where
\begin{eqnarray}
c_{f} = \frac{\Gamma\left(\frac{1}{3}\right)}{15\pi^{\frac{3}{2}}
\Gamma\left(\frac{5}{6}\right)}.
\end{eqnarray}

\section{Corrections to fermion-boson coupling\label{App:CoulombVertexCorrection}}

\begin{figure}[htbp]
\center
\includegraphics[width=3in]{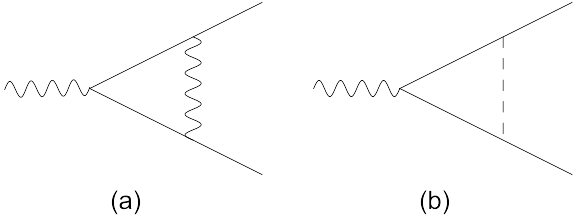}
\caption{Corrections to fermion-boson coupling due to (a) Coulomb
interaction and (b) disorder. \label{Fig:CoulombVertexCorrection}}
\end{figure}

The Feynmann diagram Fig.~\ref{Fig:CoulombVertexCorrection}(a) leads
the correction to the fermion-boson coupling
\begin{eqnarray}
\delta g^{d,t(1)} &=& -g^{3}\int'\frac{d\Omega}{2\pi}
\frac{d^3\mathbf{q}}{(2\pi)^{3}}G_{0}^{d,t}(\Omega,\mathbf{q})
G_{0}^{d,t}(\Omega,\mathbf{q})\nonumber
\\
&&\times D_{0}(\Omega,\mathbf{q}).
\label{Eq:FermionBosonCouplingCorrectionA}
\end{eqnarray}
The correction from Fig.~\ref{Fig:CoulombVertexCorrection}(b) takes
the form
\begin{eqnarray}
\delta g^{d,t(2)} = g\sum_{j=0}^{3}\Delta_{j}\int'
\frac{d^3\mathbf{k}}{(2\pi)^{3}}\Gamma_{j}
G_{0}^{d,t}(0,\mathbf{k})G_{0}^{d,t}(0,\mathbf{k})\Gamma_{j}.
\nonumber \\
\label{Eq:FermionBosonCouplingCorrectionB}
\end{eqnarray}

\begin{figure}[htbp]
\center
\includegraphics[width=3.3in]{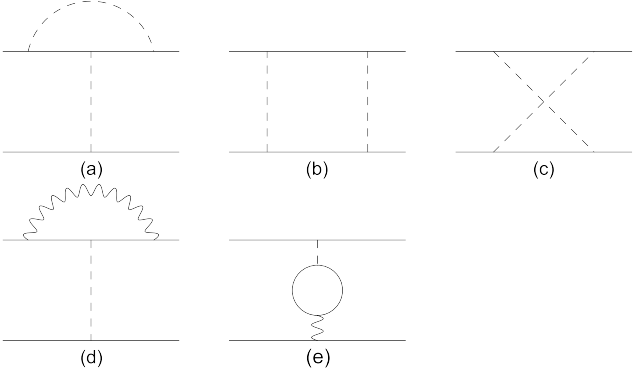}
\caption{One-loop Feynman diagrams for the corrections to the
fermion-disorder vertex. \label{Fig:VertexCorrection}}
\end{figure}

\subsection{Double-Weyl fermions \label{App:CoulombVertexCorrectionDW}}

Substituting Eq.~(\ref{Eq:FermionPropagatorDW}) into
Eqs.~(\ref{Eq:FermionBosonCouplingCorrectionA}) and
(\ref{Eq:FermionBosonCouplingCorrectionB}), we obtain the following
total corrections:
\begin{eqnarray}
\delta g^{d} = \delta g^{d(1)}+\delta g^{d(2)} =
g\sum_{j=0}^{3}\frac{\Delta_{j}}{8\pi vA}\ell.
\end{eqnarray}

\subsection{Triple-Weyl fermions \label{App:CoulombVertexCorrectionTW}}

Substituting Eq.~(\ref{Eq:FermionPropagatorTW}) into
Eqs.~(\ref{Eq:FermionBosonCouplingCorrectionA}) and
(\ref{Eq:FermionBosonCouplingCorrectionB}), we find that
\begin{eqnarray}
\delta g^{t} = \delta g^{t(1)}+\delta g^{t(2)} =
g\sum_{j=0}^{3}\frac{5}{4}\frac{\Delta_{j}c_{f}}{vB^{\frac{2}{3}}
\Lambda^{\frac{1}{3}}}\ell.
\end{eqnarray}

\section{Corrections to fermion-disorder vertex
\label{App:DisVertexCorrection}}

The correction to fermion-disorder vertex, shown by the Feynman
diagrams in Fig.~\ref{Fig:VertexCorrection}(a), is
\begin{eqnarray}
\delta \Delta_{i}^{d,t(1)} \Gamma_{i} &=&
2\Delta_{i}\sum_{j=0}^{3}\Delta_{j}\int'
\frac{d^3\mathbf{k}}{(2\pi)^{3}}\Gamma_{j}
G_{0}^{d,t}(\omega,\mathbf{k})
\Gamma_{i}\nonumber
\\
&&\times G_{0}^{d,t}(\omega,\mathbf{k})\Gamma_{j}.
\label{Eq:VertexCorrectionA}
\end{eqnarray}
The Figs.~\ref{Fig:VertexCorrection}(b) and (c) induce the
correction
\begin{eqnarray}
V^{d,t(2)+(3)} = \sum_{i=0}^{3}\sum_{i\le j\le3}V_{ij}^{d,t(2)+(3)},
\label{Eq:VertexCorrectionB1}
\end{eqnarray}
where
\begin{eqnarray}
V_{ij}^{d,t(2)+(3)} &=&
2\Delta_{i}\Delta_{j}\int'\frac{d^3\mathbf{k}}{(2\pi)^{3}}
\left(\psi_{a}^{\dag}\Gamma_{i} G_{0}^{d,t}(0,\mathbf{k})
\Gamma_{j}\psi_{a}\right)\nonumber
\\
&&\times\left\{\psi_{b}^{\dag}
\left[\Gamma_{j}G_{0}^{d,t}(0,\mathbf{k})\Gamma_{i}\right.\right.\nonumber
\\
&&\left.\left. +
\Gamma_{i}G_{0}^{d,t}(0,-\mathbf{k})\Gamma_{j}\right]\psi_{b}\right\}.
\label{Eq:VertexCorrectionB2}
\end{eqnarray}
There are ten choices for the values of $i$ and $j$. The correction
to fermion-disorder vertices due to Coulomb interaction, as
displayed in Fig.~\ref{Fig:VertexCorrection}(d), can be written as
\begin{eqnarray}
V_{i}^{d,t(4)} &=& -2\Delta_{i}g^{2}\int' \frac{d\Omega}{2\pi}
\frac{d^3\mathbf{q}}{(2\pi)^{3}} G_{0}^{d,t}(\Omega,\mathbf{q})
\Gamma_{i}G_{0}^{d,t}(\Omega,\mathbf{q})\nonumber
\\
&&\times D_{0}(\Omega,\mathbf{q}).
\label{Eq:VertexCorrectionC}
\end{eqnarray}
Fig.~\ref{Fig:VertexCorrection}(e) gives rise to
\begin{eqnarray}
\delta \Delta_{i}^{d,t(5)} &=& 2\Delta_{i}g^{2}\int'
\frac{d\omega}{2\pi}\frac{d^3\mathbf{k}}{(2\pi)^{3}}
\mathrm{Tr}\left[G_{0}^{d,t}(\omega,\mathbf{k})
\Gamma_{i}\right.\nonumber
\\
&&\left.\times G_{0}^{d,t}(\omega+\Omega,\mathbf{k}+\mathbf{q})\right]
D_{0}(\Omega,\mathbf{q}). \label{Eq:VertexCorrectionD}
\end{eqnarray}

\subsection{Double-Weyl fermions \label{App:DisVertexCorrectionDW}}

We substitute Eqs.~(\ref{Eq:FermionPropagatorDW}) and
(\ref{Eq:BosonPropagator}) into
Eqs.~(\ref{Eq:VertexCorrectionA})-(\ref{Eq:VertexCorrectionD}), and
derive the fermion-disorder vertex corrections:
\begin{eqnarray}
\delta\Delta_{0}^{d} &=& \left[\frac{1}{4\pi vA}
\left(\Delta_{0}^{2}+\frac{3}{2}\Delta_{0}\Delta_{1} +
\frac{3}{2}\Delta_{0}\Delta_{2} +
\Delta_{0}\Delta_{3}\right.\right.\nonumber
\\
&&\left.\left.+\Delta_{1}\Delta_{2}\right) -
2\Delta_{0}\left(C_{\bot}^{d} +
\frac{C_{z}^{d}}{\eta}\right)\right]\ell,
\\
\delta\Delta_{1}^{d} &=& \left[\frac{1}{8\pi vA}\left(\Delta_{0}^{2}
+ \Delta_{2}^{2} + \Delta_{3}^{2} +
2\Delta_{0}\Delta_{2}+2\Delta_{1}\Delta_{2}\right.\right.\nonumber
\\
&&\left.+\Delta_{1}\Delta_{3}-\Delta_{0}\Delta_{1}\right) +
2\Delta_{1}C_{4}^{d}\Big]\ell,
\\
\delta\Delta_{2}^{d} &=& \left[\frac{1}{8\pi vA}\left(\Delta_{0}^{2}
+ \Delta_{1}^{2} + \Delta_{3}^{2} +
2\Delta_{0}\Delta_{1}+2\Delta_{1}\Delta_{2} \right.\right.\nonumber
\\
&&\left.+
\Delta_{2}\Delta_{3}-\Delta_{0}\Delta_{2}\right) +
2\Delta_{2}C_{4}^{d}\Big]\ell,
\\
\delta\Delta_{3}^{d} &=& \left[\frac{1}{8\pi vA}
\left(\Delta_{1}\Delta_{3} + \Delta_{2}\Delta_{3}\right)
+2\Delta_{3}C_{3}^{d}\right]\ell,
\end{eqnarray}
where
\begin{eqnarray}
C_{4}^{d} = \frac{\alpha}{4\pi}\int_{0}^{+\infty}d\delta
\frac{\delta^{2}+2}{\left(1+\delta^2\right)} \frac{1}{\delta\left(1
+ \delta^2\right)^{\frac{1}{2}} + \zeta^{d}}.
\end{eqnarray}

\subsection{Triple-Weyl fermions\label{App:DisVertexCorrectionTW}}

Substituting Eqs.~(\ref{Eq:FermionPropagatorTW}) and
(\ref{Eq:BosonPropagator}) into
Eqs.~(\ref{Eq:VertexCorrectionA})-(\ref{Eq:VertexCorrectionD}), we
finally get
\begin{eqnarray}
\delta\Delta_{0}^{t} &=& \left[\left(\frac{5}{2}\Delta_{0}^{2} +
\frac{5}{2}\Delta_{0}\Delta_{1} + \frac{5}{2}\Delta_{0}\Delta_{2} +
\frac{5}{2}\Delta_{0}\Delta_{3}\right.\right.\nonumber
\\
&&\left. + 3\Delta_{1}\Delta_{2} +
\frac{1}{2}\Delta_{1}\Delta_{3}+
\frac{1}{2}\Delta_{2}\Delta_{3}\right)
\frac{c_{f}}{vB^{\frac{2}{3}}\Lambda^{\frac{1}{3}}}\nonumber
\\
&&\left.-2\Delta_{0}\left(C_{\bot}^{t} +
\frac{C_{z}^{t}}{\eta}\right)\right]\ell,
\\
\delta\Delta_{1}^{t} &=& \left[\left(-\frac{3}{2}
\Delta_{1}\Delta_{0}-\frac{3}{2}\Delta_{1}^{2} +
\frac{3}{2}\Delta_{1}\Delta_{2} +
\frac{3}{2}\Delta_{1}\Delta_{3}\right.\right.\nonumber
\\
&&\left. +
3\Delta_{0}\Delta_{2}+\frac{1}{2}\Delta_{0}\Delta_{3}\right)
\frac{c_{f}}{vB^{\frac{2}{3}}\Lambda^{\frac{1}{3}}}\nonumber
\\
&&+2\Delta_{1}C_{4}^{t}\Big]\ell,
\\
\delta\Delta_{2}^{t} &=& \left[\left(-\frac{3}{2}\Delta_{2}
\Delta_{0}+\frac{3}{2}\Delta_{2}\Delta_{1}-\frac{3}{2}\Delta_{2}^{2}
+ \frac{3}{2}\Delta_{2}\Delta_{3}\right.\right.\nonumber
\\
&&\left.+3\Delta_{0}\Delta_{1} + \frac{1}{2}\Delta_{0}
\Delta_{3}\right)\frac{c_{f}}{vB^{\frac{2}{3}}
\Lambda^{\frac{1}{3}}}\nonumber
\\
&&+2\Delta_{2}C_{4}^{t}\Big]\ell,
\\
\delta\Delta_{3}^{t} &=& \left[\frac{1}{2}\left(\Delta_{3}
\Delta_{0} - \Delta_{3}\Delta_{1}-\Delta_{3}\Delta_{2} +
\Delta_{3}^{2}+\Delta_{0}\Delta_{1} \right.\right.\nonumber
\\
&&\left.\left.+\Delta_{0}\Delta_{2}\right)
\frac{c_{f}}{vB^{\frac{2}{3}}\Lambda^{\frac{1}{3}}}\ell +
2\Delta_{3}C_{3}^{t}\right]\ell,
\end{eqnarray}
where
\begin{eqnarray}
C_{4}^{t} &=& \frac{\alpha}{6\pi}\int_{0}^{+\infty} d\delta \frac{2
+ \delta^{2}}{\delta^{\frac{1}{3}}\left(1 +
\delta^{2}\right)^{\frac{5}{6}}}\nonumber
\\
&&\times\frac{1}{\delta^{\frac{2}{3}}
\left(1+\delta^{2}\right)^{\frac{2}{3}}+ \zeta^{t}}.
\end{eqnarray}

\section{Derivation of the RG equations \label{App:RGEquations}}

The action of free multi-Weyl fermions is
\begin{eqnarray}
S_{\psi_{d,t}}^{0} &=& \int\frac{d\omega}{2\pi}
\frac{d^3\mathbf{k}}{(2\pi)^{3}}\psi_{d,t}^{\dag}(\omega,\mathbf{k})
\left[-i\omega+H_{d,t}(\mathbf{k})\right]\nonumber
\\
&&\times\psi_{d,t}(\omega,\mathbf{k}).
\end{eqnarray}
Including the self-energy corrections to the above action leads to
\begin{eqnarray}
S_{\psi_{d,t}} &=& \int\frac{d\omega}{2\pi}
\frac{d^3\mathbf{k}}{(2\pi)^{3}}\psi_{d,t}^{\dag} \left[-i\omega +
H_{d,t}(\mathbf{k}) - \Sigma_{C}^{d,t}\right.\nonumber
\\
&&\left. -
\Sigma_{\mathrm{dis}}^{d,t}\right]\psi_{d,t}.
\end{eqnarray}
It can be further written as
\begin{eqnarray}
S_{\psi_{d}} &\approx& \int\frac{d\omega}{2\pi}
\frac{d^3\mathbf{k}}{(2\pi)^{3}}\psi_{d}^{\dag}(\omega,\mathbf{k})
\Big\{i\omega e^{\left(C_{1}^{d} +
\sum_{j=0}^{3}\frac{\Delta_{j}}{8\pi vA}\right)\ell}\nonumber
\\
&& -A\left[d_{1}(\mathbf{k})\sigma_{1} +
d_{2}(\mathbf{k})\sigma_{2}\right] e^{C_{2}^{d}\ell} -
vk_{z}e^{C_{3}^{d}\ell}\sigma_{3}\Big\}\nonumber
\\
&&\times\psi_{d}(\omega,\mathbf{k}),
\end{eqnarray}
and
\begin{eqnarray}
S_{\psi_{t}} &\approx& \int\frac{d\omega}{2\pi}
\frac{d^3\mathbf{k}}{(2\pi)^{3}}\psi_{t}^{\dag}(\omega,\mathbf{k})
\Big\{i\omega e^{\sum_{j=0}^{3}\frac{5}{2} \frac{c_{f}
\Delta_{j}}{vB^{\frac{2}{3}} \Lambda^{\frac{1}{3}}}\ell}\nonumber
\\
&&- B\left[g_{1}(\mathbf{k})\sigma_{1} + g_{2}
(\mathbf{k})\sigma_{2}\right] e^{C_{2}^{t}\ell}\nonumber
\\
&&-vk_{z}\sigma_{3} e^{C_{3}^{t}\ell}\Big\}\psi_{t}(\omega,\mathbf{k}),
\end{eqnarray}
respectively. For double-Weyl fermions, we make the following
re-scaling transformations
\begin{eqnarray}
\omega&=&\omega'e^{-\ell}, \label{Eq:TransformationOmegaDW}
\\
k_{x}&=&k_{x}'e^{-\frac{\ell}{2}}, \label{Eq:TransformationkxDW}
\\
k_{y}&=&k_{y}'e^{-\frac{\ell}{2}}, \label{Eq:TransformationkyDW}
\\
k_{z}&=&k_{z}'e^{-\ell}, \label{Eq:TransformationkzDW}
\\
\psi_{d} &=& \psi_{d}'e^{\left(2-\frac{1}{2}
\sum_{j=0}^{3}\frac{\Delta_{j}}{8\pi vA}\right)\ell},
\label{Eq:TransformationpsiDW}
\\
A&=&A'e^{\left(-C_{2}^{d}+\sum_{j=0}^{3} \frac{\Delta_{j}}{8\pi
vA}\right)\ell}, \label{Eq:TransformationADW}
\\
v&=&v'e^{\left(-C_{3}^{d}+\sum_{j=0}^{3} \frac{\Delta_{j}}{8\pi
vA}\right)\ell}, \label{Eq:TransformationvDW}
\end{eqnarray}
and, for triple-Weyl fermions, we make the following transformations
\begin{eqnarray}
\omega&=&\omega'e^{-\ell}, \label{Eq:TransformationOmegaTW}
\\
k_{x}&=&k_{x}'e^{-\frac{\ell}{3}}, \label{Eq:TransformationkxTW}
\\
k_{y}&=&k_{y}'e^{-\frac{\ell}{3}}, \label{Eq:TransformationkyTW}
\\
k_{z}&=&k_{z}'e^{-\ell}, \label{Eq:TransformationkzTW}
\\
\psi_{t}&=&\psi_{t}' e^{\left(\frac{11}{6} -
\frac{1}{2}\sum_{j=0}^{3}\frac{5}{4}\frac{c_{f}\Delta_{j}}{
vB^{\frac{2}{3}}\Lambda^{\frac{1}{3}}}\right)\ell},\label{Eq:TransformationpsiTW}
\\
B&=&B'e^{\left(-C_{2}^{t}+\sum_{j=0}^{3}\frac{5}{4}\frac{c_{f}\Delta_{j}}{
vB^{\frac{2}{3}}\Lambda^{\frac{1}{3}}}\right)\ell}, \label{Eq:TransformationATW}
\\
v&=&v'e^{\left(-C_{3}^{t}+\sum_{j=0}^{3}\frac{5}{4}\frac{c_{f}\Delta_{j}}{
vB^{\frac{2}{3}}\Lambda^{\frac{1}{3}}}\right)\ell}.\label{Eq:TransformationvTW}
\end{eqnarray}
Now the action of fermions becomes
\begin{eqnarray}
S_{\psi_{d,t}'} &=&
\int\frac{d\omega'}{2\pi}\frac{d^3\mathbf{k}'}{(2\pi)^{3}}
\psi_{d,t}'^{\dag}(\omega',\mathbf{k}')\left[-i\omega' +
H'_{d,t}(\mathbf{k}')\right]\nonumber
\\
&&\times\psi_{d,t}'(\omega',\mathbf{k}'),
\end{eqnarray}
which has the same form as the free action.

The free action of bosonic field $\phi$ is
\begin{eqnarray}
S_{\phi}^{0} = \int\frac{d\omega}{2\pi}
\frac{d^3\mathbf{k}}{(2\pi)^{3}}\phi(\omega,\mathbf{k})
\left(k_{x}^{2}+k_{y}^{2}+\eta k_{z}^{2}\right)
\phi(\omega,\mathbf{k}).
\end{eqnarray}
After including the self-energy corrections, we modify the free
action to
\begin{eqnarray}
S_{\phi} &=& \int\frac{d\omega}{2\pi}
\frac{d^{3}\mathbf{k}}{(2\pi)^{3}}\phi(\omega,\mathbf{k})
\left(k_{x}^{2}+k_{y}^{2} + \eta k_{z}^{2}+\Pi^{d,t}(0,\mathbf{k})
\right)\nonumber
\\
&&\times\phi(\omega,\mathbf{k})\nonumber
\\
&\approx&\int\frac{d\omega}{2\pi} \frac{d^3\mathbf{k}}{(2\pi)^{3}}
\phi(\omega,\mathbf{k}) \left[k_{\bot}^{2}e^{C_{\bot}^{d,t}\ell} +
\left(\eta + C_{z}^{d,t}\ell\right) k_{z}^{2}\right]\nonumber
\\
&&\times\phi(\omega,\mathbf{k}).
\end{eqnarray}
We then employ the transformations
Eqs.~(\ref{Eq:TransformationOmegaDW})-(\ref{Eq:TransformationkzDW}),
and
\begin{eqnarray}
\phi = \phi'e^{\frac{1}{2}\left(4 - C_{\bot}^{d}\right)\ell}
\label{Eq:TransformationPhiDW}
\end{eqnarray}
for double-WSM, We then employ the transformations
Eqs.~(\ref{Eq:TransformationOmegaTW})-(\ref{Eq:TransformationkzTW}) and
\begin{eqnarray}
\phi = \phi'e^{\frac{1}{2}\left(\frac{10}{3} -
C_{\bot}^{t}\right)\ell} \label{Eq:TransformationPhiTW}
\end{eqnarray}
for triple-WSM. The action of $\phi'$ can now be rewritten as
\begin{eqnarray}
S_{\phi'} &\approx& \int\frac{d\omega'}{2\pi}
\frac{d^{3}\mathbf{k}'}{(2\pi)^{3}}\phi'(\omega',\mathbf{k}')
\left\{k_{\bot}'^{2} + \left[\eta-\eta\left(1 +
C_{\bot}^{d}\right)\ell\right.\right.\nonumber
\\
&&\left.\left. +
C_{z}^{d}\ell\right]k_{z}'^{2} \right\}\phi'(\omega',\mathbf{k}')
\end{eqnarray}
for the double-WSM and
\begin{eqnarray}
S_{\phi'} &\approx& \int\frac{d\omega'}{2\pi}
\frac{d^{3}\mathbf{k}'}{(2\pi)^{3}}\phi'(\omega',\mathbf{k}')
\left\{k_{\bot}'^{2}+\left[\eta-\eta\left(\frac{4}{3} +
C_{\bot}^{t}\right)\ell\right.\right.\nonumber
\\
&&\left.\left.+C_{z}^{t}\ell\right]k_{z}'^{2}
\right\}\phi'(\omega',\mathbf{k}').
\end{eqnarray}
for the triple-WSM. It is convenient to define
\begin{eqnarray}
\eta' = \eta-\eta\left(1+C_{\bot}^{d}\right)\ell +
C_{z}^{d}\ell,\label{Eq:TransformationEtaDW}
\end{eqnarray}
for the double-WSM and
\begin{eqnarray}
\eta' = \eta - \eta\left(\frac{4}{3}+C_{\bot}^{t}\right)\ell +
C_{z}^{t}\ell \label{Eq:TransformationEtaTW}
\end{eqnarray}
for the triple-WSM. The action for boson sector then becomes
\begin{eqnarray}
S_{\phi'} &=& \int\frac{d\omega'}{2\pi}
\frac{d^{3}\mathbf{k}'}{(2\pi)^{3}} \phi'(\omega',\mathbf{k}')
\left(k_{\bot}'^{2}+\eta'k_{z}'^{2}\right)\nonumber
\\
&&\times\phi'(\omega',\mathbf{k}'),
\end{eqnarray}
which is formally the same as the free action.

The bare action of the fermion-boson coupling is
\begin{eqnarray}
S_{\psi\phi}^{0} &=& g\int\frac{d\omega_{1}}{2\pi}
\frac{d^3\mathbf{k}_{1}}{(2\pi)^{3}}\frac{d\omega_{2}}{2\pi}
\frac{d^3\mathbf{k}_{2}}{(2\pi)^{3}}
\psi_{d,t}^{\dag}(\omega_{1},\mathbf{k}_{1})
\psi_{d,t}(\omega_{2},\mathbf{k}_{2})\nonumber
\\
&&\times\phi(\omega_{1} -
\omega_{2},\mathbf{k}_{1} - \mathbf{k}_{2}).
\end{eqnarray}
Including the corrections to this interaction vertex, we rewrite the
above action as
\begin{eqnarray}
S_{\psi\phi} &=& \left(g + \delta g^{d,t}\right)
\int\frac{d\omega_{1}}{2\pi}\frac{d^3\mathbf{k}_{1}}{(2\pi)^{3}}
\frac{d\omega_{2}}{2\pi} \frac{d^3\mathbf{k}_{2}}{(2\pi)^{3}}
\psi_{d,t}^{\dag}(\omega_{1},\mathbf{k}_{1})\nonumber
\\
&&\times\psi_{d,t}(\omega_{2},\mathbf{k}_{2})\phi(\omega_{1} -
\omega_{2},\mathbf{k}_{1} - \mathbf{k}_{2}).
\end{eqnarray}
Making use of the transformations
(\ref{Eq:TransformationOmegaDW})-(\ref{Eq:TransformationpsiDW}) and
(\ref{Eq:TransformationPhiDW}) for the double-WSM, and
(\ref{Eq:TransformationOmegaTW})-(\ref{Eq:TransformationpsiTW}) and
(\ref{Eq:TransformationPhiTW}) for the triple-WSM, we find that the
coupling parameter should transform as follows:
\begin{eqnarray}
g' = ge^{-\frac{C_{\bot}^{d,t}}{2}\ell}.
\label{Eq:Transformationg}
\end{eqnarray}
We then write the action of fermion-boson coupling in the form
\begin{eqnarray}
S_{\psi'\phi'} &=& g'\int\frac{d\omega_{1}'}{2\pi}
\frac{d^3\mathbf{k}_{1}'}{(2\pi)^{3}}\frac{d\omega_{2}'}{2\pi}
\frac{d^3\mathbf{k}_{2}'}{(2\pi)^{3}}
\psi_{d,t}'^{\dag}(\omega_{1}',\mathbf{k}_{1}')\nonumber
\\
&&\times\psi_{d,t}'(\omega_{2}',\mathbf{k}_{2}')\phi'(\omega_{1}' -
\omega_{2}',\mathbf{k}_{1}' - \mathbf{k}_{2}),
\end{eqnarray}
which recovers the form of the bare action.

Including the leading order corrections to the fermion-disorder
vertex leads to the following action term
\begin{eqnarray}
S_{\mathrm{dis}}^{F} &=& \sum_{i=0}^{3}\frac{\left(\Delta_{i} +
\delta\Delta_{i}^{d,t}\right)}{2} \int\frac{d\omega_1
d\omega_2d^{3}\mathbf{k}_{1} d^{3}\mathbf{k}_{2}
d^{3}\mathbf{k}_{3}}{(2\pi)^8}\nonumber
\\
&&\times\psi^\dagger_{a}(\omega_1,\mathbf{k}_1)\Gamma_{i}
\psi_{a}(\omega_1,\mathbf{k}_2)
\psi^\dagger_{b}(\omega_2,\mathbf{k}_3)\Gamma_{i}\nonumber
\\
&&\times\psi_{b}(\omega_2,-\mathbf{k}_1-\mathbf{k}_2-\mathbf{k}_3).
\end{eqnarray}
By employing the re-scaling transformations
(\ref{Eq:TransformationOmegaDW})-(\ref{Eq:TransformationpsiDW}) for
the double-WSM and
(\ref{Eq:TransformationOmegaTW})-(\ref{Eq:TransformationpsiTW}) for
the triple-WSM, we obtain
\begin{eqnarray}
S_{\mathrm{dis}}^{F} &\approx& \sum_{i=0}^{3}\frac{\left(\Delta_{i} +
\delta\Delta_{i}^{d} - 2\Delta_{i}\sum_{j=0}^{3}
\frac{\Delta_{j}}{8\pi vA}\ell\right)}{2}\nonumber
\\
&&\times\int\frac{d\omega_1'
d\omega_2'd^{3}\mathbf{k}_{1}' d^{3}\mathbf{k}_{2}'
d^{3}\mathbf{k}_{3}'}{(2\pi)^8}
\psi_{a}'^\dagger(\omega_1',\mathbf{k}_1')
\Gamma_{i}\nonumber
\\
&&\times\psi'_{a}(\omega_1',\mathbf{k}_2')
\psi'^\dagger_{b}(\omega_2',\mathbf{k}_3')\Gamma_{i} \nonumber \\
&&\times\psi'_{b}(\omega_2',-\mathbf{k}_1'-\mathbf{k}_2'-\mathbf{k}_3'),
\end{eqnarray}
or
\begin{eqnarray}
S_{\mathrm{dis}}^{F} &\approx& \sum_{i=0}^{3}\frac{1}{2}\Bigg[\Delta_{i}
\left(1+\frac{1}{3}\ell\right)+\delta\Delta_{i}^{t}\nonumber
\\
&&-2\Delta_{i}
\sum_{j=0}^{3}\frac{5}{4}\frac{c_{f}\Delta_{j}}{vB^{\frac{2}{3}}
\Lambda^{\frac{1}{3}}}\ell\Bigg]\int\frac{d\omega_1'
d\omega_2'd^{3}\mathbf{k}_{1}' d^{3}\mathbf{k}_{2}'
d^{3}\mathbf{k}_{3}'}{(2\pi)^8}\nonumber
\\
&&\times\psi'^\dagger_{a}(\omega_1',\mathbf{k}_1')\Gamma_{i}
\psi'_{a}(\omega_1',\mathbf{k}_2')
\psi'^\dagger_{b}(\omega_2',\mathbf{k}_3')\Gamma_{i}\nonumber
\\
&&\times\psi'_{b}(\omega_2',-\mathbf{k}_1'-\mathbf{k}_2'-\mathbf{k}_3'),
\end{eqnarray}
which apply to the double- and triple-WSMs, respectively. We define
\begin{eqnarray}
\Delta_{i}' = \Delta_{i}+\delta\Delta_{i}^{d} -
2\Delta_{i}\sum_{j=0}^{3}\frac{\Delta_{j}}{8\pi vA}\ell
\label{Eq:TransforamtionDeltaiDW}
\end{eqnarray}
for the double-WSM and
\begin{eqnarray}
\Delta_{i}' = \Delta_{i}\left(1+\frac{1}{3}\ell\right) +
\delta\Delta_{i}^{t}-2\Delta_{i}\sum_{j=0}^{3}\frac{5}{4}
\frac{c_{f}\Delta_{j}}{ vB^{\frac{2}{3}}\Lambda^{\frac{1}{3}}}\ell
\label{Eq:TransforamtionDeltaiTW}
\end{eqnarray}
for the triple-WSM, and finally have
\begin{eqnarray}
S_{\mathrm{dis}}^{F} &=& \sum_{i=0}^{3}\frac{\Delta_{i}'}{2}
\int\frac{d\omega_1' d\omega_2'd^{3}\mathbf{k}_{1}'
d^{3}\mathbf{k}_{2}'d^{3}\mathbf{k}_{3}'}{(2\pi)^8}
\psi'^\dagger_{a}(\omega_1',\mathbf{k}_1')\nonumber
\\
&&\times\Gamma_{i}
\psi_{a}'(\omega_1',\mathbf{k}_2')
\psi'^\dagger_{b}(\omega_2',\mathbf{k}_3')\Gamma_{i}\nonumber
\\
&&\times\psi_{b}'(\omega_2',-\mathbf{k}_1'-\mathbf{k}_2'-\mathbf{k}_3').
\end{eqnarray}

\subsection{Double-Weyl fermions}

According to
Eqs.~(\ref{Eq:TransformationpsiDW})-(\ref{Eq:TransformationvDW}),
(\ref{Eq:TransformationEtaDW}), (\ref{Eq:Transformationg}),
(\ref{Eq:TransforamtionDeltaiDW}), we can get the RG equations for
$\Delta_{i}$ and other parameters:
\begin{eqnarray}
\frac{dZ_{f}}{d\ell} &=& -\frac{1}{2}
\sum_{j=0}^{3}\Delta_{j}Z_{f},\label{Eq:RGEqZfDW}
\\
\frac{dA}{d\ell}&=&\left(C_{2}^{d} -
\frac{1}{2}\sum_{j=0}^{3}\Delta_{j}\right)A,\label{Eq:RGEqADW}
\\
\frac{dv}{d\ell}&=&\left(C_{3}^{d} -
\frac{1}{2}\sum_{j=0}^{3}\Delta_{j}\right)v,\label{Eq:RGEqVFDW}
\\
\frac{d\alpha}{d\ell}&=& \left(-C_{\bot}^{d}-C_{3}^{d} +
\frac{1}{2}\sum_{j=0}^{3}\Delta_{j}\right)\alpha,
\label{Eq:RGEqAlphaDW}
\\
\frac{d\beta^{d}}{d\ell} &=& \left(1+C_{3}^{d}-C_{2}^{d}  -
\beta^{d}\right)\beta^{d}, \label{Eq:RGEqBetaDW}
\\
\frac{d\eta}{d\ell}&=&\left(-1-C_{\bot}^{d}+\beta^{d}\right)\eta,
\\
\frac{dg}{d\ell}&=&-\frac{C_{\bot}^{d}}{2}g,
\\
\frac{d\Delta_{0}}{d\ell} &=& \left(\Delta_{0}^{2}+\frac{3}{2}
\Delta_{0}\Delta_{1}+\frac{3}{2}\Delta_{0}\Delta_{2}
+\Delta_{0}\Delta_{3}+\Delta_{1}\Delta_{2}\right)\nonumber
\\
&&-\Delta_{0}\left(C_{2}^{d}+C_{3}^{d}+2C_{\bot}^{d} +
2\beta^{d}\right), \label{Eq:RGEqDelta0DW}
\\
\frac{d\Delta_{1}}{d\ell} &=& \frac{1}{2}\left(\Delta_{0}^{2}
+\Delta_{2}^{2}+\Delta_{3}^{2}+2\Delta_{0}\Delta_{2} +
2\Delta_{1}\Delta_{2}+\Delta_{1}\Delta_{3}\right.\nonumber
\\
&&\left.-\Delta_{0}\Delta_{1}\right)
+ \Delta_{1} \left(2C_{4}^{d}-C_{2}^{d}-C_{3}^{d}\right),
\label{Eq:RGEqDelta1DW}
\\
\frac{d\Delta_{2}}{d\ell} &=& \frac{1}{2}\left(\Delta_{0}^{2} +
\Delta_{1}^{2}+\Delta_{3}^{2}+2\Delta_{0}\Delta_{1}+2\Delta_{1}
\Delta_{2}+\Delta_{2}\Delta_{3}\right.\nonumber
\\
&&\left.-\Delta_{0}\Delta_{2}\right)+\Delta_{2}
\left(2C_{4}^{d}-C_{2}^{d}-C_{3}^{d}\right), \label{Eq:RGEqDelta2DW}
\\
\frac{d\Delta_{3}}{d\ell}
&=&\frac{1}{2}\left(\Delta_{1}\Delta_{3}+\Delta_{2}\Delta_{3}\right)
-\Delta_{3}\left(C_{2}^{d}-C_{3}^{d}\right), \label{Eq:RGEqDelta3DW}
\end{eqnarray}
where
\begin{eqnarray}
\alpha=\frac{g^{2}}{4\pi v},\qquad
\beta^{d}=\frac{C_{z}^{d}}{\eta}=\frac{g^{2}}{64\pi v}
\frac{v^{2}}{A\Lambda\eta}.
\end{eqnarray}
In the derivation, we have used the following re-definition
\begin{eqnarray}
\frac{\Delta_{i}}{4\pi vA}\rightarrow\Delta_{i}.
\end{eqnarray}

\subsection{Triple-Weyl fermions}

Using
Eqs.~(\ref{Eq:TransformationpsiTW})-(\ref{Eq:TransformationvTW}),
(\ref{Eq:TransformationEtaTW}), (\ref{Eq:Transformationg}),
Eq.~(\ref{Eq:TransforamtionDeltaiTW}), we finally obtain the RG
equations
\begin{eqnarray}
\frac{dZ_{f}}{d\ell}&=&-\frac{5}{4}\sum_{j=0}^{3}\Delta_{j}Z_{f},
\\
\frac{dB}{d\ell}&=&\left(C_{2}^{t}-\frac{5}{4}
\sum_{j=0}^{3}\Delta_{j}\right)B,
\\
\frac{dv}{d\ell}&=&\left(C_{3}^{t}-\frac{5}{4}
\sum_{j=0}^{3}\Delta_{j}\right)v,
\\
\frac{d\alpha}{d\ell} &=& \left(-C_{\bot}^{t}-C_{3}^{t} +
\frac{5}{4}\sum_{j=0}^{3}\Delta_{j}\right)\alpha,
\\
\frac{d\beta^{t}}{d\ell}
&=&\left(\frac{4}{3}+C_{3}^{t}-\frac{2}{3}C_{2}^{t}
-\beta^{t}-\frac{5}{12}\sum_{j=0}^{3}\Delta_{j}\right)\beta^{t},
\\
\frac{d\eta}{dl}&=&\left(-\frac{4}{3}-C_{\bot}^{t}+\beta^{t}\right)\eta,
\\
\frac{dg}{dl}&=&-\frac{C_{\bot}^{t}}{2}g,
\\
\frac{d\Delta_{0}}{d\ell} &=& \frac{1}{3}\Delta_{0}
+\left(\frac{25}{12}\Delta_{0}^{2} +
\frac{25}{12}\Delta_{0}\Delta_{1} +
\frac{25}{12}\Delta_{0}\Delta_{2} \right.\nonumber
\\
&&\left.+\frac{25}{12}\Delta_{0}\Delta_{3} +
3\Delta_{1}\Delta_{2}+\frac{1}{2}\Delta_{1}\Delta_{3}+
\frac{1}{2}\Delta_{2}\Delta_{3}\right)\nonumber
\\
&&-\Delta_{0}\left(\frac{2}{3}C_{2}^{t} +
C_{3}^{t}+2C_{\bot}^{t}+2\beta^{t}\right),
\\
\frac{d\Delta_{1}}{d\ell} &=&\frac{1}{3}\Delta_{1}+
\left(-\frac{23}{12}\Delta_{1}\Delta_{0} - \frac{23}{12}
\Delta_{1}^{2} + \frac{13}{12}\Delta_{1}\Delta_{2}\right.\nonumber
\\
&&\left.+\frac{13}{12}\Delta_{1}\Delta_{3} + 3\Delta_{0}\Delta_{2} +
\frac{1}{2}\Delta_{0}\Delta_{3}\right)\nonumber
\\
&&+\Delta_{1}\left(2C_{4}^{t}-\frac{2}{3}C_{2}^{t}-C_{3}^{t}\right),
\\
\frac{d\Delta_{2}}{d\ell}
&=&\frac{1}{3}\Delta_{2}+\left(-\frac{23}{12}\Delta_{2}\Delta_{0} +
\frac{13}{12}\Delta_{2}\Delta_{1}-\frac{23}{12}\Delta_{2}^{2}\right.\nonumber
\\
&&\left.+\frac{13}{12}\Delta_{2}\Delta_{3}+3\Delta_{0}\Delta_{1} +
\frac{1}{2}\Delta_{0}\Delta_{3}\right)\nonumber
\\
&&+\Delta_{2}\left(2C_{4}^{t}-\frac{2}{3}C_{2}^{t}-C_{3}^{t}\right),
\\
\frac{d\Delta_{3}}{d\ell} &=& \frac{1}{3}\Delta_{3}
+\left(\frac{1}{12}\Delta_{3}\Delta_{0} -
\frac{11}{12}\Delta_{3}\Delta_{1} -
\frac{11}{12}\Delta_{3}\Delta_{2}\right.\nonumber
\\
&&\left.+\frac{1}{12}\Delta_{3}^{2} +
\Delta_{0}\Delta_{1}+\Delta_{0}\Delta_{2}\right)\nonumber
\\
&&-\Delta_{3}\left(\frac{2}{3}C_{2}^{t}-C_{3}^{t}\right),
\end{eqnarray}
where
\begin{eqnarray}
\beta^{t} = \frac{C_{z}^{t}}{\eta} =
\frac{\Gamma\left(\frac{1}{3}\right)}{120\pi^{\frac{3}{2}}
\Gamma\left(\frac{5}{6}\right)}\frac{g^{2}v}{B^{\frac{2}{3}}
\Lambda^{\frac{4}{3}}\eta}.
\end{eqnarray}
The re-definition
\begin{eqnarray}
\frac{c_{f}\Delta_{i}}{vB^{\frac{2}{3}}\Lambda^{\frac{1}{3}}}
\rightarrow \Delta_{i}
\end{eqnarray}
has been employed in the derivation of RG equations.

\section{Expression of $C_{Li}$ \label{App:CLiLRDisorder}}

The expressions of $C_{Li}$ with $i=1, 2, 3$ are given by
\begin{eqnarray}
C_{L1}&=&\frac{1}{6\pi^{2}c_{f}}\int_{0}^{+\infty}d\delta
\frac{\left(1+\delta^{2}\right)^{\frac{1}{6}}}{\delta^{\frac{1}{3}}}\nonumber
\\
&&\times \frac{1}{\delta^{\frac{2}{3}}
\left(1+\delta^{2}\right)^{\frac{2}{3}}+\zeta^{t}},
\\
C_{L2}&=&-\frac{1}{6\pi^{2}c_{f}}\int_{0}^{+\infty}d\delta
\frac{1}{\delta^{\frac{1}{3}}\left(1+\delta^{2}\right)^{\frac{5}{6}}}\nonumber
\\
&&\times\left[-18\delta^{2}+1 + 45 \frac{\delta^{4}}{\left(1 +
\delta^{2}\right)}-27\frac{\delta^{6}}{\left(1+\delta^{2}\right)^{2}}\right]\nonumber
\\
&&\times\frac{1}{\delta^{\frac{2}{3}}
\left(1+\delta^{2}\right)^{\frac{2}{3}}+\zeta^{t}},
\\
C_{L3}&=&-\frac{1}{6\pi^{2}c_{f}}\int_{0}^{+\infty} d\delta
\frac{\delta^{2}-1}{\delta^{\frac{1}{3}}\left(1 +
\delta^{2}\right)^{\frac{5}{6}}}\nonumber
\\
&&\times\frac{1}{\delta^{\frac{2}{3}}\left(1 +
\delta^{2}\right)^{\frac{2}{3}}+ \zeta^{t}}.
\end{eqnarray}

\end{document}